\documentclass[A4paper,11pt]{article}
\def\spacingset#1{\renewcommand{\baselinestretch}%
{#1}\small\normalsize} \spacingset{1.5}

\usepackage[colorlinks,citecolor=blue,urlcolor=blue]{hyperref}

\usepackage{graphicx} 
\usepackage{amsmath,amsthm}
\usepackage{natbib}
\usepackage{booktabs} 
\usepackage{longtable}

\usepackage{lscape}
\usepackage{rotating}
\usepackage{hyperref}

\usepackage{savesym}
\usepackage{moreverb}
\savesymbol{bigtimes}
\restoresymbol{TXF}{bigtimes}
\usepackage{longtable}
\usepackage{subcaption}
\usepackage{prodint}
\usepackage{listings}

\usepackage{caption}

\begin{document}

\begin{center}
    {\LARGE \bf An Overview and Recent Developments in the Analysis of Multistate Processes} \\[1.5em]
    
    {\large Malka Gorfine} \\
    \textit{Department of Statistics and Operations Research, Tel Aviv University, Israel} \\[1em]

    {\large Richard J. Cook} \\
    \textit{Department of Statistics and Actuarial Science, University of Waterloo, Canada} \\[1em]

    {\large Per Kragh Andersen} \\
    \textit{Section of Biostatistics, University of Copenhagen, Denmark} \\[1em]

    {\large Terry M. Therneau} \\
    \textit{Division of Biomedical Statistics and Informatics, Mayo Clinic, New York, USA} \\[1em]

    {\large Pierre Joly} \\
    \textit{Inserm, ISPED, Bordeaux Populations Health Research Center, University of Bordeaux, France} \\[1em]

    {\large Hein Putter} \\
    \textit{Department of Biomedical Data Sciences, Leiden University Medical Center, Leiden, The Netherlands} \\[1em]

    {\large Maja Pohar Perme} \\
    \textit{Department of Biostatistics and Medical Informatics, Medical Faculty, University of Ljubljana, Slovenia} \\[1em]

    {\large Michal Abrahamowicz} \\
    \textit{Department of Epidemiology, Biostatistics and Occupational Health, McGill University, Montreal, Quebec, Canada} \\[1em]

    \textbf{On Behalf of Topic Group 8 ``Survival Analysis" of the STRATOS Initiative}
\end{center}

\newpage

\abstract{
Multistate models offer a powerful framework for studying disease processes and can be used to formulate intensity-based and more descriptive marginal regression models.
They also represent a natural foundation for the construction of joint models for disease processes and dynamic marker processes, as well as joint models incorporating random 
censoring and intermittent observation times.  
This article reviews the ways multistate models can be formed and fitted to life history data.
Recent work on pseudo-values and the incorporation of random effects to model dependence on the process history and between-process heterogeneity are also discussed.
The software available to facilitate such analyses is listed.  
}



\section{Introduction\label{section1}}
\subsection{Background}
Long-term cohort studies offer an excellent source of data for research on the onset and progression of chronic disease processes. 
When interest lies in the time to a particular event, insights are often gained from use of survival analysis methods. 
Yet, in many settings individuals may experience several types of events (i.e. myocardial infarctions, bleeds, non-fatal stroke, and death in cardiovascular research) and interest may 
lie in studying their occurrence and the relationship between these event times. 
Multistate models offer a versatile framework for studying such processes through the analysis of  transition rates between different health states. 
The objective of this paper is to provide a review of the statistical models and methodology for multistate analysis and discuss the formulation, estimation, and interpretation of alternative models, as well as to illustrate their application. 
The paper follows our first guidance paper on intensity-based models for time to event analyses \citep{andersen2021}. We hope this overview  will promote more widespread and informed use of multistate models in research on human health and advance the general aims of the STRATOS Initiative \citep{sauerbrei2014}.

Diseases involving distinct stages can be naturally characterized using multistate models.
The states may represent different stages of a  progressive condition such as hepatitis \citep{sweeting2010}, 
the presence or absence of symptoms in episodic conditions such as chronic bronchitis \cite{grossman1998},
different phases of a response to treatment in cancer clinical trials \citep{gelber1990},
or the course of a COVID-19 infection as individuals move between moderate, severe, and critical disease states in hospital, 
become discharged, or die \citep{roimi2021development,rossman2021hospital}.
In other settings, multistate models have  been used to characterize the passage of insects through successive developmental stages of their life cycle \citep{munholland1991}, the formation and dissolution of marriages \citep{espenshade1982}, or changes in the employment status of individuals in the workforce \citep{zarghami2024}.
Careful modeling of  process dynamics can yield valuable scientific insights regarding the natural disease course, risk factors for disease onset and progression, 
and intervention effects. 
Multistate models also offer a useful framework for the specification of joint models for time-dependent covariate processes and disease processes of interest; such models can improve 
understanding of the dynamic relationship between two or more processes. 

\begin{figure}
\centering
\includegraphics[width=0.9\textwidth]{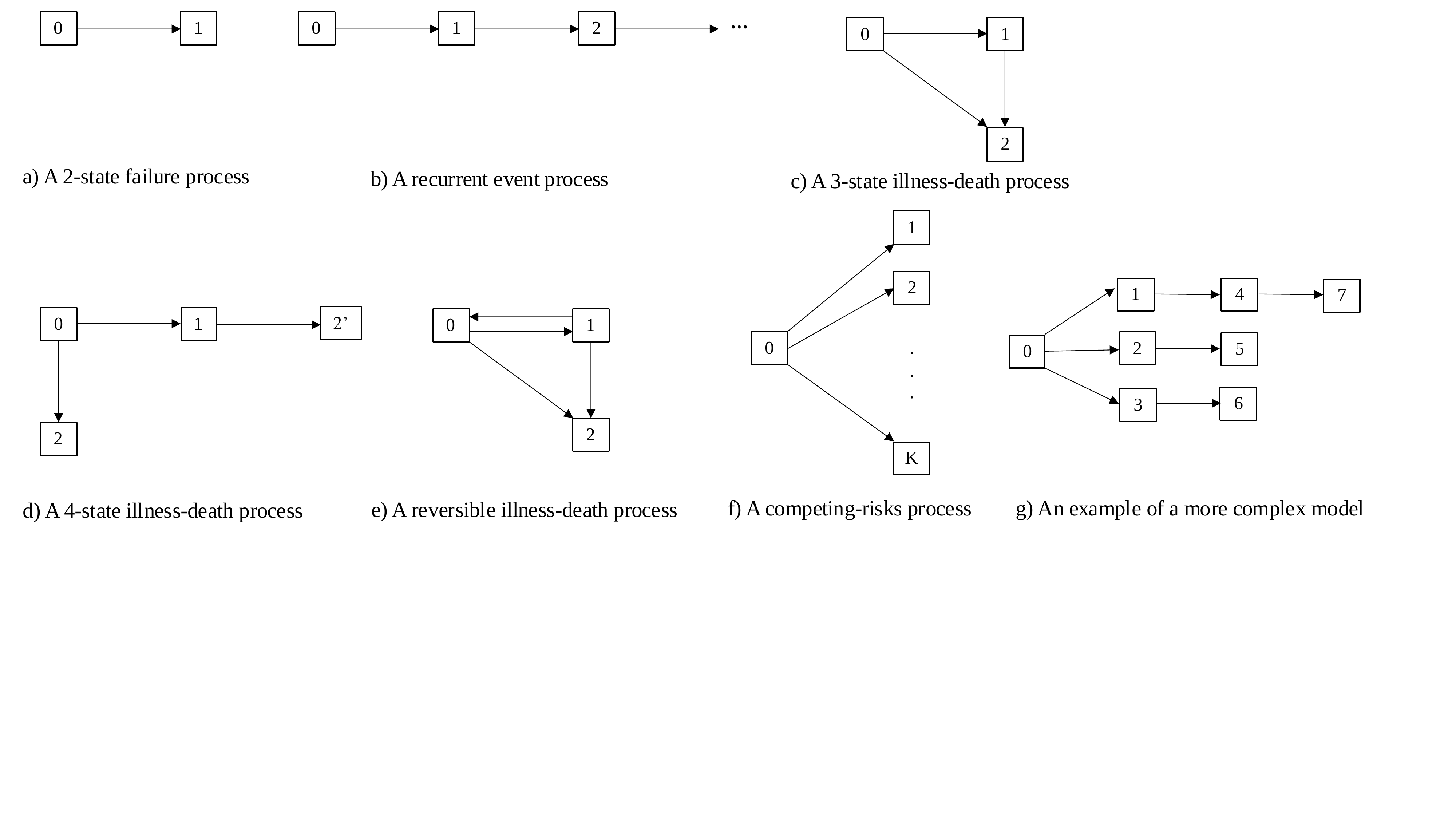}
\caption{Some common processes represented as multistate models \label{fig1} }
\end{figure}

The formation of multistate models begins with the specification of a set of states ${\cal S}$ representing meaningfully different conditions of the process. 
These typically correspond to different states of health or different stages of a disease process. 
The set ${\cal S}$ typically has a finite number of elements we label with integers $\{0, 1, 2,\ldots,K\}$, say, but processes with a countably number of states can
also be modeled (e.g. recurrent event processes) \citep{cook-lawless-book2018}.
There is often a subset of states that can be entered directly from a given state with this subset determined from the context. 
Fig. \ref{fig1} contains some illustrative state-space diagrams representing some of the rich variety of processes that can be analyzed using multistate models. The arrows in a state-space diagram represent the transitions that can be made directly.
When processes can terminate, the set of absorbing states from which individuals cannot exit is denoted by 
 ${\cal A}$.

Fig. \ref{fig1} a) is a two-state diagram representing a simple failure process with state 0 a transient state and state 1 an absorbing state entered upon failure.
Fig. \ref{fig1} b) is a multistate diagram representing a recurrent event process \citep{cook2007} where the state labels correspond to the cumulative number of events experienced; 
there is no absorbing state represented here.
Fig. \ref{fig1} c) is a conventional illness-death model which  is helpful to describe many  processes in public health such as disease
onset or progression when death may also occur. 
For  disease incidence, state 0 may represent an initial healthy state, state 1 is entered upon disease onset and state 2 is entered upon death which may happen 
in disease-free individuals (as a $0 \rightarrow 2$ transition), or after the development of the disease (corresponding to a $1 \rightarrow 2$ transition).
Fig. \ref{fig1} d) is an alternative representation of an illness death model with ${\cal A} = \{2, 2'\}$ where the different absorbing states convey whether deaths occurred in disease-free 
 or diseased individuals. 
This distinction between \ref{fig1} c) and d) may seem trivial but the distinction between event-free death and death following an intermediate event can enable more informative simple 
descriptive analyses.
Fig. \ref{fig1} e) is a reversible illness-death model which, because the $1 \rightarrow 0$ transition is allowed, is suitable for chronic obstructive pulmonary disease (COPD) where symptoms may develop  or resolve over time \citep{sood2016spirometric}.
Finally, the setting of $K+1$ states with $K$ competing events is depicted in Fig. \ref{fig1} f). 
Much more elaborate state spaces can be defined to represent more complex processes, for example Fig. \ref{fig1} g), and the principles and methods we discuss next apply in such settings. In addition, a time origin must be specified. This can sometimes be obvious as in the case of survival data where it is the birth of an individual, but sometimes its specification can be challenging. Discrete-time models can be adopted but we focus on continuous-time models in which event times take values on the positive real line.

\subsection{Examples of Multistate Survival Data}
The following publicly available datasets are used throughout the text to illustrate various approaches to multistate analysis. Codes are available in the Supplementary Material file.

\subsubsection{Recurrence and Death in Colon Cancer} 
In a clinical trial conducted in the 1980s to investigate the use of levamisole and fluorouracil as adjuvant therapies for resected colon carcinoma \citep{moertel1990levamisole}, a total of 929 patients with stage C disease were randomly assigned to one of three groups: observation, levamisole alone, or levamisole combined with fluorouracil. The time to cancer recurrence and  survival time were considered key outcomes. We consider the illness-death process of Fig. \ref{fig1} d), where state 1 is entered upon cancer recurrence, state 2 represents recurrence-free death, and state $2'$ represents post-recurrence death. State 0 corresponds to the initial state at the time of treatment assignment which we set as the time origin ($t=0$). The dataset is available in the \texttt{R} package \texttt{survival} \citep{survival-package}.
The use of levamisole has no effect, so we combine observation and levamisole arms as the control and code treatment as
5FU+Lev (1) vs Control (0).
A total of 468 individuals experienced recurrence and  452 individuals were observed to die; most death (414) occurred following recurrence.
We revisit this example in Section 2.2 to illustrate estimation of cumulative transition intensities and state occupancy probabilities over time.
In Section 2.4 we illustrate findings from intensity-based regression modeling. 

\subsubsection{Joint Damage in Psoriatic Arthritis}

A second study we consider involves the progression of joint damage in 305 individuals with psoriatic arthritis. 
The dataset from the University of Toronto Psoriatic Arthritis Cohort is available in the {\tt msm R} package \citep{msm}, where four states are defined to represent varying levels of disease severity: mild, moderate, severe, and very severe impairment. 
Specifically, state 0 corresponds to 0 damaged joints, state 1 to  1-4 damaged joints, state 2 to 5-9 damaged joints, and state 3 to  $10$ or more  damaged joints.
This is a progressive process (i.e., transitions are irreversible) with transitions occurring in continuous time, but information on the number of damaged joints is only collected during periodic radiographic examinations,  
scheduled to take place every two years. 
The actual timing of assessments varies considerably across individuals, resulting in random visit times.
Fig. \ref{fig:psoriatic} a) illustrates the recruitment times and subsequent times of radiographic examination for five patients.
In Section 4, we consider Fig.\ref{fig1}b), with  four states and two covariates—based on the number of effusions and the sedimentation rate—and model their effect on transitions to more advanced states of damage. An effusion is a swelling of tissue around a joint due to a build-up of fluid which signals the disease is in an active phase; the count of the number of effusions is a measure of disease activity at the joint or patient level. Erythrocyte sedimentation rate is a blood marker reflecting the systemic level of inflammation.

\subsubsection{Rotterdam Tumor Bank Data}
Data from the Rotterdam tumor bank, which includes 1,546 breast cancer patients who had node-positive disease and underwent a tumor removal surgery between the years 1978–1993, are available in the \texttt{survival R} package \citep{survival-package}. We take the date of surgery for tumor removal 
as the time of origin for Fig. \ref{fig1} d); date of relapse,
 date of relapse-free death, and date of post-relapse death   are the respective entry times to states 1, 2 and $2'$.  Prognostic baseline variables are age at surgery, menopausal status, tumor size, tumor grade, number of positive lymph nodes, levels of estrogen and progesterone receptors in the initial biopsy, hormonal therapy, and chemotherapy. Of the 1546 patients, 924 experienced a relapse of the disease
(63\%), 106 died without evidence of relapse (7\%), and 771 patients died after a relapse (79\% of the patients who showed a relapse of the cancer). This dataset is used to demonstrate the frailty-based methods discussed in Section 5.1.

\subsection{Overview of the Paper}

In Section 2.1 we define intensity functions, which serves as the building blocks for full models of multistate processes. 
Different classes  of intensity functions are introduced,  distinguished by the time scale governing risk. We then discuss computation of transition probability matrices for Markov processes and consider functionals of intensities, which are often the target of inference.
Nonparametric estimation methods for a single sample with processes subject to right censoring, are presented in Section 2.2. In Section 2.3, we discuss the formulation of multiplicative intensity-based models to study the effect of fixed or time-varying covariates on transition rates. In Section 2.4 we derive the likelihood based on a joint model for time-varying covariates, a censoring time, and the process of interest. Section 3  focuses on regression models aimed at estimating state occupancy probabilities using so-called pseudo-values. Section 4 addresses scenarios where processes operating in continuous time are observed intermittently, typically during clinic visits; see the example in Section 1.2.2. Section 5 explores the use of random effects in modeling multistate disease processes. Section 6 provides a review of statistical packages and functions available for multistate analysis. Concluding remarks and suggestions for future research are presented in Section 7. The appendix includes a glossary of the notation used throughout the paper for ease of reference.

\section{Methodology}
\subsection{Notation and Foundations}

We let $t=0$ represent the time origin of the process and  assume that processes are observed from their onset unless otherwise specified.
The state occupied at time $t \ge 0$ by an individual is represented by $Z(t)$ and in what follows we assume $\Pr(Z(0)=0)=1$ unless otherwise stated.
The multistate process is denoted by $\{Z(s), 0 \le s\}$ 
with the history $H(t) = \{Z(s), 0\le s < t\}$ at time $t>0$ representing a record of the number, types, and times of transitions over $[0, t)$. 
For progressive processes, wherein each state can be entered at most once, it is convenient to define $T_k$ as the entry time to state $k$. 
If states are recurrent then one can let $T_k^{(r)}$ denote the $r$th entry time to state $k$, $r=1,\ldots $, $k=0, 1, \ldots$.
Counting process notation offers another helpful representation of multistate processes wherein $N_{kl}(t)$ records the number of $k \rightarrow l$ transitions over $[0,t]$; 
in this case  $N_{kl}(t)$ increments by one at $T_l^{(r)}$ if state $l$ is entered from state $k$ at $T_l^{(r)}$. 

Intensity functions are the fundamental building blocks of multistate processes, with the $k\rightarrow l$ intensity function given by
\begin{equation}\label{intensity}
\lambda_{kl}(t|H(t))= \lim_{\Delta t \downarrow 0} 
\displaystyle\frac{1}{\Delta t}{ \Pr\left(Z((t + \Delta t)^-)=l | Z(t^-)=k, H(t)\right)} \, , \,\,\, k \neq l \, ,
\end{equation}
where $t^-$ denotes  infinitesimal amount of time before $t$. Note that (\ref{intensity})
 is a very general function and model specification involves explicit representation of how the history affects the risk of transitions.
Markov models have the property that the intensity functions do not depend on the history of the process beyond the state occupied at $t^-$, so
$\lambda_{kl}(t|H(t))= \lambda_{kl}(t|Z(t^-)=k)=\lambda_{kl}(t)$. 
For semi-Markov processes the intensity depends on the time since entry to state $k$, so we  write 
$\lambda_{kl}(t|H(t))= \lambda_{kl}(B(t)|Z(t^-)=k)$ where $B(t)= t-t_k^{(n_k(t))}$ with
$n_k(t) = \sum_{j\ge 0} N_{jk}(t^-)$ the total number of times state $k$ was entered over $[0, t)$ and 
$t_k^{(n_k(t))}$  the time that state $k$ was most recently entered at $t>0$.
A given process may involve intensities for some transitions which have a Markov form and some which have a semi-Markov form.
Moreover, a given intensity function may involve a hybrid time-scale.
For example, in chronic obstructive pulmonary disease, individuals may have recurrent periods with exacerbations of symptoms, which may arise with increasing frequency as the disease duration increases;  
the time to resolution of the exacerbations may also tend to increase with increasing  time since disease onset.
The intensity in such cases can involve specification of a basic time scale and incorporate dependence on other aspects of time via regression. For instance, one can adopt models of the form $\lambda_{kl}(t|H(t))= \lambda_{kl}\left(B(t)|Z(t^-)=k\right) \exp\left( g(t)^{\sf T} \gamma\right)$ where $g(t)$ is a function of time and $\gamma$ is a parameter that characterizes the dependence on $t$. 

The continuous-time Markov model is a canonical model that warrants special attention.
As noted earlier, the intensity $\lambda_{kl}(t|H(t))= \lambda_{kl}(t)$ for such processes, and we define the cumulative transition intensity as 
$\Lambda_{kl}(t)= \int_0^t d\Lambda_{kl}(s)$ where $d\Lambda_{kl}(s)=\lambda_{kl}(s)ds$.
If the state space ${\cal S} = \{0, 1, \ldots, K\}$, then a $(K+1) \times (K+1)$ transition intensity matrix $d{\mathbf \Lambda}(s)$ can be formed with off-diagonal entries 
$d \Lambda_{kl}(s)$ and  the $(k,k)$ diagonal entry $-\sum_{l\ne k} d \Lambda_{kl}(s)$.
If $\mathbf{I}$ is a $(K+1) \times (K+1)$ identity matrix, by product integration \citep{cook-lawless-book2018} one can obtain
\begin{equation} 
\label{productintegration}
\mathbf{P}(s,t)=\Prodi _{(s,t]}\left\{\mathbf{I}+d \mathbf{\Lambda}(u) \right\}
\end{equation}
where 
$\mathbf{P}(s,t)$ is a $(K+1)\times (K+1)$ transition probabilities matrix with $(k,l)$ entry $p_{kl}(s,t) = \Pr(Z(t)=l|Z(s)=k)$.
When $s=0$, we can compute the probability that a particular state is occupied at  time $t>0$, denoted as $p_k(t) =\Pr(Z(t)=k|Z(0)=0)$. This result forms the basis for defining a wide range of marginal features, including:
\begin{itemize}
\item[(i)] The probability that the process is in one of a set ${\cal S}^\dagger$ of states is given by $\sum_{k \in {\cal S}^\dagger} p_k(t)$, ${\cal S}^\dagger \subseteq {\cal S}$. If ${\cal S}^\dagger = {\cal A}$,
this corresponds to the probability that the process has terminated by time $t>0$.
\item[(ii)] The expected total sojourn time in state $k$ is $\mu_k = \int_0^\infty p_k(u)du$. A \textit{restricted mean sojourn time} over the interval $(0,\tau]$ can be 
defined by specifying a finite upper limit of integration, expressed as
$\mu_k(\tau) = \int_0^\tau p_k(u)du$.
\item[(iii)] For progressive processes, it defines the cumulative incidence function for state $k$, 
given by
$F_k(t) = \Pr(T_k \leq t) = \sum_{j \in {\cal S}_k} p_j(t)$ where ${\cal S}_k \subseteq {\cal S}$ includes state $k$ and any states that can be entered following a sojourn in state $k$.
This is the probability of having entered state $k$ by time $t$.  
\end{itemize}

We next discuss nonparametric estimation of $\mathbf{P}(s, t)$.
This is arguably the main focus of most real-world applications and is illustrated in Section 2.2. 

\subsection{Censored Data and Descriptive Functionals for Multistate Processes}

Suppose a sample of $n$ independent individual processes is observed from $t=0$ up to a maximum of  $\tau$, where $\tau$ is a fixed administrative censoring time.
Let $i$ denote the index for individuals in the sample, $i=1,\ldots, n$. 
Since individuals may be lost to follow-up, let $C^*_i$ be a random right-censoring time  for individual $i$. Each individual is thus observed continuously  over the interval $[0, C_i]$, where 
$C_i= \min (C^*_i, \tau)$. 

Defining the at-risk process and counting processes for events is more involved in a general multistate setting compared to the single failure time setting.  The  function $Y_i(t) = I(t \leq C_i)$ indicates whether the process for individual $i$ is under observation (i.e. is uncensored) at  time $t>0$. 
If $N_{ikl}(t)$ is the total number of $k \rightarrow l$ transitions over $[0,t]$ for process $i$, $\Delta N_{ikl}(t) = N_{ikl}((t+\Delta t)^-)-N_{ikl}(t^-)$ represents the number  of $k \rightarrow l$ transitions for individual $i$  over $[t, t+\Delta t)$, and 
$dN_{ikl}(t) = \lim_{\Delta t \downarrow 0} \Delta N_{ikl}(t)$ is an indicator that they experienced a $k \rightarrow l$ transition at $t$.
To distinguish the underlying counting process and the process observed under this censoring scheme,
let $Y_{ik}(t) = I(Z_i(t)=k)$ indicate that state $k$ is occupied at $t$ by individual $i$,  and $\bar{Y}_{ik}(t) = Y_i(t) Y_{ik}(t^-)$ indicates a transition of individual $i$ out of state $k$ may be observed at time $t$. 
Then, let 
$d \bar{N}_{ikl}(t) = \bar{Y}_{ik}(t) \, d N_{ikl}(t)$
indicate that a $k \rightarrow l$ transition is recorded for process $i$ at time $t$, and $\bar{N}_{ikl}(t) = \int_0^t \bar{Y}_{ik}(s) \, d N_{ikl}(s)$ denote the total 
number of $k \rightarrow l$ transitions observed for process $i$  over the interval $[0, t]$. If $k \in {\cal A}$,  then $dN_{ikl}(t)$ is zero for all $t>0$ and all $l \in \mathcal{S}$, since absorbing states cannot be exited.

A natural estimator of $d\Lambda_{kl}(t)$ is given by
\begin{equation}
\label{na1}
d\widehat{\Lambda}_{kl}(t)=  \displaystyle\frac{d\bar{N}_{\cdot kl}(t)}{\bar{Y}_{\cdot k}(t)}
\end{equation}
where $d\bar{N}_{\cdot kl}(t) = \sum_{i=1}^n \bar{Y}_{ik}(t) dN_{ikl}(t)$ is the total number of $k \rightarrow l$ transitions at time $t$ observed in the sample, and 
$\bar{Y}_{\cdot k}(t) =\sum_{i=1}^n \bar{Y}_{ik}(t)$  is the total number of individuals at risk of a $k \rightarrow l$ transition in the sample at time $t$.
The Nelson-Aalen estimator of $\Lambda_{kl}(t)$ is then
\begin{equation}
\label{na}
\widehat{\Lambda}_{kl}(t) = \displaystyle\int_0^t d \widehat{\Lambda}_{kl}(u) \; ,
\end{equation}
where the integral is a Stieltjes integral representation of a discrete sum over the distinct $k \rightarrow l$ transition times over $(0,t]$. It is apparent from  (\ref{na1})
that the integrand will be zero except at times when $k \rightarrow l$ transitions are observed.

Replacing the unknown quantities in the right-hand side of (\ref{productintegration}) with the estimates given by (\ref{na}), and carrying out product 
integration, gives the Aalen-Johansen estimator \citep{aalen1978} 
\begin{equation}
\label{AJ}
\widehat{\mathbf{P}}(s, t) = \Prodi_{(s,t]}\left\{ \mathbf{I}+ d \widehat{\mathbf \Lambda}(u) \right\}
\end{equation}
where $\widehat{\mathbf{\Lambda}}(u)$ is the matrix of estimated cumulative transition intensities obtained by replacing $\Lambda_{kl}(t)$ with the Nelson-Aalen estimator.
If processes are observed from $s=0$ and $\Pr(Z(0)=0)=1$, the top row of $\widehat{\mathbf{P}}(s, t)$ contains the Aalen-Johansen estimator of the state $k$ occupancy probability, $p_{k}(t)$, $k=0,1,\ldots, K$.
This in turn enables estimation of the functionals (i) - (iii) along with many others.
Importantly, while this nonparametric estimator is motivated by the Markov assumption, the estimates of $p_k(t)$ are robust and valid for non-Markov processes provided 
censoring is completely independent \cite{aalen2001covariate}.

\subsubsection{The Colon Cancer Study Revisited, I}
To illustrate, we consider the colon cancer data set, introduced in Section 1.2.1. The code is available in Section S1 of the Supplementary Material file. The data frame is structured in the ``counting process'' format, making it suitable for analyzing in terms of a broad class of multistate processes. In this format, the follow-up period for each individual is divided into intervals during which the individual is at risk of transitioning from one state to any other possible state. The presence of the $\bar{Y}_{ik}(t)$ term in the denominator of Eq. (\ref{na1}) necessitates tracking when individuals occupy different states, specifically when they are at risk of transitioning out of state $k$. 

\begin{figure}
\centering
\includegraphics[width=0.9\textwidth]{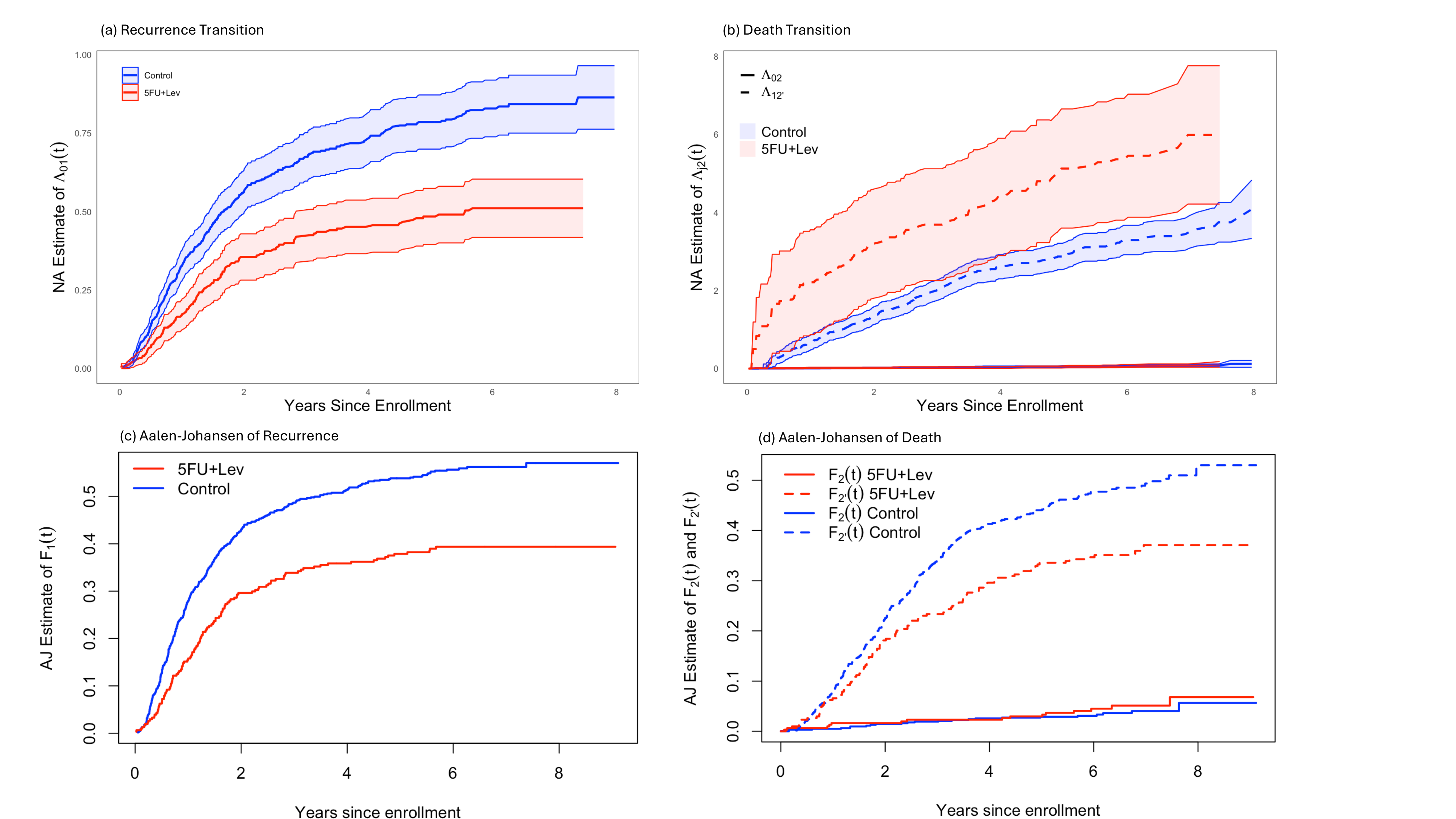}
\caption{Colon cancer study: Nelson-Aalen (NA) estimates of cumulative transition intensities by treatment groups along with pointwise 95\% confidence intervals, and Aalen-Johansen (AJ) estimates of the cumulative incidence functions for recurrence and recurrence-free death and post-recurrence death.}\label{fig:colon-all} 
\end{figure}

Fig. \ref{fig:colon-all} (a) shows the Nelson-Aalen estimates of the cumulative transition intensity for recurrence ($0 \rightarrow 1$) along with pointwise 95\% confidence intervals. The slope of these estimates reflects the magnitude and trend in the of recurrence intensity among individuals who  are alive and recurrence-free (i.e., those in state 0). The control group shows a roughly constant intensity over the first two years, followed by a lower intensity. The 5FU+Lev group shows a lower initial intensity, leveling off after the initial two years.

Fig. \ref{fig:colon-all} (b) presents the Nelson-Aalen estimates for transitions into death states ($ 0 \rightarrow 2$ and $1 \rightarrow 2'$) and pointwise 95\% confidence intervals. 
The very small estimated probabilities  reflect that relatively few individuals make transitions directly from state 0 to 2 for either treatment group. Among individuals who experience early recurrence, however, those individuals receiving 5FU+Lev have an elevated intensity for death; see the sharp increase within the first six months. The estimate for the control group appears more linear.  
However, for state $k=1$ the denominator of Eq. (\ref{na1}) may be very small in the first few months and this may be an artefact --  the slopes of $1\rightarrow 2'$ are roughly similar after this initial six months period. Fig. \ref{fig:colon-all} (b) suggests a slightly elevated risk of death in the 5FU+Lev group compared to the control arm. The overall reduction in post-recurrence mortality in the 5FU+Lev group is primarily due to a lower risk of recurrence. The separation in the Nelson-Aalen estimates is largely driven by differences observed during the early follow-up period, when the number of subjects at risk is relatively small.

If $T_1$ is the time of entry into state 1, then the cumulative incidence function for recurrence, $F_1(t)= \Pr(T_1 \leq t)$,  can be expressed as the probability of having entered state 1 by time $t$,
given by $F_1(t)=  p_1(t) + p_{2'}(t)$.
Likewise,  the cumulative incidence function for recurrence-free death is defined as $F_2(t) = p_2(t)$, while the probability of death following recurrence is given by $F_{2'}(t) = p_{2'}(t)$.
Note that these are all sub-distribution functions because they do not approach 1 as $t \rightarrow \infty$ due to the presence of competing risks. Such functions, however, have the appealing feature of being interpretable as probabilities.

The Aalen-Johansen estimates of $F_1(t)$ are shown in  Fig. \ref{fig:colon-all} (c) for each treatment group. It is evident that the 5FU+Lev group has an appreciably lower risk of recurrence. Fig.
\ref{fig:colon-all} (d) presents the estimates for the cumulative incidence functions of death-related events. Evidently, the risk of recurrence-free death is low in both treatment groups, and 5FU+Lev is associated with a reduced risk of post-recurrence death.

\subsection{Intensity-Based Regression Models}

If interest lies in assessing the relationship between time-varying covariates and the multistate process, this can be studied through intensity-based regression models \citep{andersen1993statistical,aalen2008survival,beyersmann2011competing,cook-lawless-book2018}.
Let $X(t)$ represent a $p \times 1$ covariate at $t$ and $\{X(s), 0\le s \}$ denote the covariate process. 
If ${\cal H}(t) = \{Z(s), X(s), 0\le s < t\}$ is the expanded history including information on the covariate path, the intensity function can be modified by replacing $H(t)$ with ${\cal H}(t)$ in Eq. (\ref{intensity}).Intensity-based regression models aim to characterize how the instantaneous risk of a $k\rightarrow l$ transition intensity depends on features of the covariate process. The most common form is the multiplicative model \citep{andersen1993statistical, cook-lawless-book2018, beyersmann2011competing, therneau2024multi} where
\begin{equation}
\label{intensity2}
\lambda_{kl}(t|{\cal H}(t))= \lambda_{kl}(t|H(t)) \exp(X(t)^{\sf T} \beta_{kl}) \, .
\end{equation}
If $\lambda_{kl}(t|H(t))= \lambda_{kl}(t)$, this is a modulated Markov model, where the covariate process modulates the baseline Markov intensity.
Likewise, if $\lambda_{kl}(t|H(t))=\lambda_{kl}(B(t))$ in (\ref{intensity2}), this corresponds to a modulated semi-Markov model \citep{cook-lawless-book2018}. When covariates are time fixed, conditional on the covariates $X$, these models reduce to Markov and semi-Markov models, respectively.
For Markov processes given covariates $X$, a $(K+1) \times (K+1)$ transition probability matrix ${\mathbf P}(s,t|X)$ can be defined
with $(k,l)$ entry $\Pr(Z(t)=l|Z(s)=k, X)$. This can be estimated by stratification if the covariates are discrete, or by fitting regression models like
(\ref{intensity2}) and applying product integration as in (\ref{productintegration}). 
Covariate effects can also be expressed as having additive effects on the process intensities \cite{aalen2001covariate}.
For processes which are non-Markov it is harder to compute transition probabilities \cite{dabrowska1995estimation,spitoni2012estimation}.

The stochastic nature of the multistate process is fully specified by the set of transition intensities  \citep{andersen1993statistical, aalen2008survival}.
When covariates are time-varying, joint models for the covariate and disease processes are often useful. We next discuss constructing likelihoods in the presence of time-dependent covariates and random censoring, and address likelihood construction under intermittent observation in Section 3.

\subsection{Likelihood for Intensity-Based Models with Time-Dependent Covariates} \label{sec1.4}

When processes involve time-dependent covariates and random censoring, it is important to recognize that these are random processes which play a role in the data generation. Here, we provide a discussion of the likelihood construction in this setting \cite{cook-lawless-book2018}.

Let $C_i^*(t)  = I(C_i^* \leq t)$ be the indicator of whether random censoring occurred by time $t$,
and denote the corresponding counting process as $\{C_i^*(s), 0 \le s\}$. Usually, information on the covariates ceases when the multistate process enters an absorbing state or the process is censored and we assume this in what follows. Hence, the at-risk process should be adjusted accordingly. Let $Y_i^\dagger(t) = I(Z_i(t) \notin {\cal A})$ indicate that the occupied state at time $t$ by  individual $i$ is a non-absorbing state, 
$\bar{Y}_i(t)=Y_i(t) Y_i^\dagger(t^-)$ equals 1 if individual $i$ may be observed to transit at time $t$.  
The vector $\bar{N}_{ik}(t)= (\bar{N}_{ikl}(t), l \ne k, l = 1, \ldots, K)^{\sf T}$ records the cumulative number of transitions from state $k$ over $(0,t]$ where it is understood that these counts will be zero for state $l$ that can not be entered directly from state $k$.  Finally,   $\bar{N}_i(t)= (\bar{N}^{\sf T}_{ik}(t), k \notin {\cal A})^{\sf T}$ be the vector of all counting processes of non-absorbing states.
We define $\Delta \bar{X}_i(t) = \bar{Y}_i(t + \Delta t)  \{ X_i( (t + \Delta t)^-) - X_i( t^- ) \}$ to represent an increment in the observed covariate vector over $[t, t + \Delta t)$. Here, $\bar{Y}_i(t + \Delta t)$ ensures that the multistate process has not yet reached an absorbing state and is not yet censored. Additionally, 
$d \bar{X}_i(t) = \lim_{\Delta t \downarrow 0} \Delta \bar{X}_i(t)$ and $\bar{X}_i(t) = \int_0^t d \bar{X}_i(s)$.
The history of the observed multistate, covariate, and random censoring processes is then denoted by
$\bar{\cal H}_i(t)= \{{Y}_i(s), \bar{N}_i(s), \bar{X}_i(s), 0 \le s < t; Z_i(0), X_i(0) \}$.
The intensity for random censoring
is then defined generally as
\begin{equation} \label{eq-cens-int-Yt-lambdaCt}
\lim_{\Delta t \downarrow 0} \frac{1}{ \Delta t}{\Pr(\Delta C_i^*(t) = 1 \mid \bar{\cal H}_i(t))} 
=  \bar{Y}_i(t) \, \lambda^c (t \mid \bar{\cal H}_i(t)) \, ,
\end{equation}
where $\Delta C_i^*(t) = C_i^*( (t + \Delta t)^{-} ) - C_i^*(t^{-})$. The term $\bar{Y}_i(t)$  in Eq. (\ref{eq-cens-int-Yt-lambdaCt}) ensures that the censoring intensity is zero once the multistate process is censored or reaches an absorbing state.

To construct the full likelihood, we consider a partition of $[0, \tau]$ defined by the points $0 = u_0 < u_1 < \cdots < u_R = \tau$.
We then consider the contributions over the sub-intervals $[u_{r-1}, u_r)$, $r = 1, \ldots, R$.
To this end we let
$\Delta \bar{X}_i(u_r) = \bar{Y}_i(u_r) \{ X_i(u_r) - X_i(u_{r-1}) \} $ represent an increment in the covariate vector  and let
$\Delta \bar{N}_i(u_r) = Y_i(u_r)  \{ N_i(u_r^{-}) - N_i(u_{r-1}^{-}) \}$ denote the number of transitions over $[u_{r-1},u_r)$.
Finally,  
$\bar{H}_i(u_r)= \{Y_i(u_s), \Delta \bar{N}_i(u_s), \Delta \bar{X}_i(u_s), s= 1, \ldots, r; Z_i(0), X_i(0) \}$ is the history of the censoring and joint multistate and covariates processes over the partition.
For interval $[u_{r-1}, u_r)$ the following contribution is made by processes $i$:
{\begin{align}
& \biggl\{\biggl[\Pr\left(\Delta \bar{X}_i(u_r) \mid \Delta \bar{N}_i(u_r),  Y_i(u_r) = 1,  \bar{H}_i(u_{r-1})\right) ~
\Pr\left(\Delta \bar{N}_i(u_r) \mid Y_i(u_r) = 1,  \bar{H}_i(u_{r-1}) \right) \biggr]^{Y_i(u_r)} \nonumber \\
& \quad \times~
\Pr\left(\Delta C_i^*(u_r) = 1 \mid \bar{H}_i(u_{r-1})\right)^{\Delta C_i^* (u_r)}    
\Pr\left(\Delta C_i^*(u_r) = 0 \mid \bar{H}_i(u_{r-1})\right)^{1-\Delta C_i^* (u_r)}    
 \biggr\}^{Y_i(u_{r-1}) I(Z_i(u_{r-1}) \not\in {\cal A})} \, .  \nonumber
\end{align}}
\noindent
Note that a likelihood contribution is  made over $[u_{r-1}, u_r)$ by an individual only 
if they have not been censored and the multistate  process is not in an absorbing state at time $u_{r-1}$. 
Second, there is a contribution related to the multistate and covariate processes only if the individual is not censored by time $u_r$. 
Third, by adopting the particular factorization here, the stochastic model for the increment in the 
covariate process is conditional not only on $\Delta \bar{X}_i(u_{r-1}), \ldots, \Delta \bar{X}_i(u_1), X_i(0)$ 
but also on $\Delta \bar{N}_i(u_r), \ldots, \Delta \bar{N}_i(u_1)$, $Z_i(0)$ and $I(C_i > u_r)$. 
This accommodates the setting in which covariates may cease to be defined when certain (usually absorbing) states are reached in the multistate process.

Under the partition $0=u_0< u_1 < \cdots < u_R=\tau$, the full likelihood based
on data of observation $i$ over $[0, \tau]$ is the product of the following three terms:
\begin{equation}\label{eq-full-msm-LX}
\prod_{r=1}^R \Pr\left(\Delta \bar{X}_i(u_r) \mid \Delta \bar{N}_i(u_r),  Y_i(u_r) = 1,  \bar{H}_i(u_{r-1})\right)^{Y_i(u_{r}) I(Z_i(u_{r-1}) \not\in {\cal A}) }
\end{equation}
pertaining to the covariate process,
\begin{equation} \label{eq-full-msm-LN}
\prod_{r=1}^R \Pr\left(\Delta \bar{N}_i(u_r) \mid Y_i(u_r) = 1,  \bar{H}_i(u_{r-1}) \right) ^{Y_i(u_{r}) I(Z_i(u_{r-1}) \not\in {\cal A}) } 
\end{equation}
\noindent
pertaining to the multistate process, and
\begin{equation*} 
\prod_{r=1}^R \Pr\left(\Delta C_i^*(u_r) \mid \bar{H}_i(u_{r-1})\right) ^{Y_i(u_{r-1}) I(Z_i(u_{r-1}) \not\in {\cal A})} 
\end{equation*}
\noindent
for the random censoring process.

The censoring and covariate processes are said to be \textit{noninformative} if there is no information to be gained about the parameters of primary interest (i.e., those indexing the multistate process) by modeling the censoring or covariate processes. 
Thus unless interest lies in joint modeling of a covariate (often termed a ``marker process'') and the multistate process, under the assumption that the censoring and covariate processes are noninformative, it is customary to restrict attention to (\ref{eq-full-msm-LN}).
This requires an intensity for the observable counting process. 
We write the probability of a contribution for a particular interval $[u_{r-1}, u_r)$ in (\ref{eq-full-msm-LN}) as
\begin{equation*}
\Pr\left(\Delta \bar{N}_i(u_r) \mid Y_i(u_r) = 1,  \bar{H}_i(u_{r-1}) \right) 
= \prod_{k=1}^K   \Pr \left(\Delta \bar{N}_{ik}(u_r) \mid Y_i(u_r) = 1,  \bar{H}_i(u_{r-1}) \right)^{Y_{ik}(u_{r-1})} \, ,
\end{equation*}
\noindent
which can be written more explicitly as
\begin{equation} \label{eq-full-msm-LN-prob-expand}
\prod_{k=1}^K \left\{ \prod_{l \neq k = 1}^K \Pr \left(\Delta \bar{N}_{ikl} (u_r) = 1 \mid Y_i(u_r) = 1,  \bar{H}_i(u_{r-1}) \right)^{\Delta \bar{N}_{ikl} (u_r) }   
\Pr \left(\Delta \bar{N}_{ik \cdot} (u_r) = 0 \mid Y_i(u_r) = 1,  \bar{H}_i(u_{r-1}) \right)^{1-\Delta \bar{N}_{ik \cdot} (u_r) } \right\}^{Y_{ik}(u_{r-1})} \, . 
\end{equation}
where $\bar{N}_{ik.}(t) = \sum_{l\neq k =1}^K \bar{N}_{ikl}(t)$.
To proceed further, it is necessary to define the intensity for the observable counting process
\begin{equation}
\lim_{\Delta t \downarrow 0} 
\frac{1}{\Delta t}{\Pr\left(\Delta \bar{N}_{ikl} (t) = 1 \mid Y_i(t) = 1, \bar{\cal H}_i(t)\right)} \, .
\label{indep-cens}
\end{equation}
\noindent
In order to express this in terms of the intensities of the process of interest, we require an additional assumption that the random censoring is
conditionally independent of the multistate process, given the history $\bar{\cal H}_i(\cdot)$ \citep{andersen1993statistical, aalen2008survival,lawless2019new}. This is often simply referred to as independent censoring.
Under this assumption the probability in the numerator of (\ref{indep-cens}) is 
$\Pr (\Delta N_{ikl} (t) = 1 | {\cal H}_i(t))$,
and we can write the intensity (\ref{indep-cens}) as
$\bar{Y}_{ik}(t) \, \lambda_{kl}(t \mid {\cal H}_i(t))$. 
Then, by expressing (\ref{eq-full-msm-LN-prob-expand}) in terms of $\lambda_{kl}(t \mid {\cal H}_i(t))$
and taking the limit as $R \rightarrow \infty$ we obtain 
\begin{equation}\label{likelihood}
L_i \propto  
\prod_{k = 1}^{K} \prod_{l \ne k = 1}^{K}  {L}_{ikl} 
\end{equation}
where 
\begin{equation}
{L}_{ikl} \propto 
\left\{ \prod\limits_{t_r \in {\cal D}_{ikl}} \lambda_{kl}\left(t_r \mid {\cal H}_i(t_r)\right) \right\} \,
\exp \biggl( - \int_0^\infty \bar{Y}_{ik}(u) \, \lambda_{kl}\left(u \mid {\cal H}_i(u)\right) \, d u \biggr) \, ,
\label{Lkl}
\end{equation} 
\noindent
with ${\cal D}_{ikl}$ being the set of $k \rightarrow l$ transition times observed over $[0,\tau]$ for observation $i$.
The likelihood contribution presented here is for individual $i$;  for a sample of $n$ independent processes, the overall likelihood will be a product of such terms, $\prod_{i=1}^n L_i$.

\subsubsection{The Colon Cancer Study Revisited, II}
\begin{figure}
\centering
\includegraphics[width=0.5\textwidth]{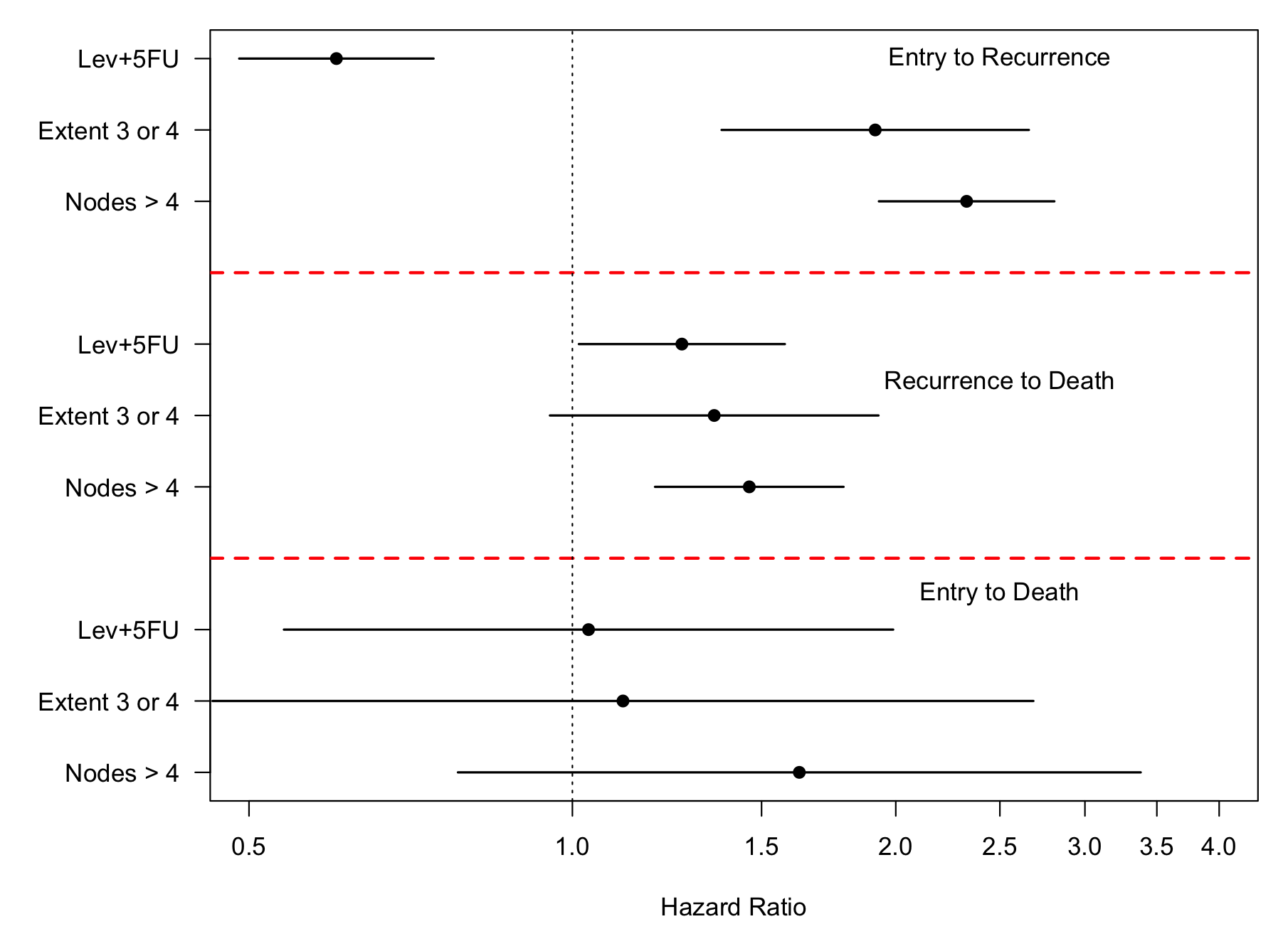}
\caption{Colon cancer study: hazard ratio \label{fig:HR95CI} and the corresponding 95\% confidence intervals from Cox regression intensity-based analyses.}
\end{figure}

To illustrate intensity-based regression modeling, we examine three risk factors in the colon cancer study:
treatment group (5FU+Lev versus control, denoted $X_1$); extent of invasion, defined as a binary variable with  submucosa or muscle (values 1 or 2 in the dataset) versus serosa  or contiguous structures (3 or 4) ($X_2$); and an indicator of more than 4 lymph nodes being involved ($X_3$).
We then apply intensity-based regression models of the form
$$
\lambda_{kl}(t|X_i) = \lambda_{kl}(t) \exp\left(X_i^{\sf T}\beta_{kl}\right)
$$
where $X_i=(X_{i1},X_{i2},X_{i3})^{\sf T}$ and $\beta_{kl}$ is a  vector of regression coefficients conveying the effect of covariates on the $k \rightarrow l$ intensity, $(k,l)\in \{(0,1), (0,2), (1,2')\}$. The code is available in Section S2 of the Supplementary Material file.
The results  in Table \ref{tbl:tab1} and Figure \ref{fig:HR95CI} show 
 that the  5FU+Lev treatment significantly reduces the rate of recurrence when controlling for the extent of disease involvement and 
nodal involvement. Likewise, individuals with more extensive disease and those with more than 4 lymph nodes involved have significantly  higher rate of recurrence when controlling for treatment.
For recurrence-free death there is no evidence of an effect of any of the risk factors.
For death following recurrence, there is a suggestion of potential harm from the treatment  when controlling for the extent of disease and nodal involvement. Alternatively, one may base analyses on pseudo-values as we discuss in Section 3.4. A more nuanced analysis of this process is warranted and would be possible with a large set of covariates.

Intensity-based regression models do not yield estimates that are robust to the omission of important covariates. In causal parlance, conditioning on
occupancy of the recurrence state creates a collider bias; see Section 8.4 of  \cite{cook-lawless-book2018}. With time-fixed covariates, an alternative way of assessing their relation to the multistate process is through stratification and computation of
expected sojourn times in different states for each stratum.

\begin{table}
\centering
\caption{Colon cancer study: Cox-regression coefficients (Est), standard errors (SE) and  hazard  ratio (HR) from intensity-based analyses.}
\label{tbl:tab1}
\begin{tabular}{@{}lrrrrr@{}} 
\toprule
Transition & Covariate & Est & SE & HR  & p-value\\
\midrule
Entry to Recurrence, $0 \rightarrow 1$ & 5FU+Lev & -0.508 & 0.106 &0.603 & $< 0.001$\\ 
& Extent 3 or 4 & 0.649 & 0.168 &1.914  & $< 0.001$\\
& Nodes $>4$ & 0.845 &  0.096 & 2.328 & $< 0.001$\\
\midrule
Recurrence to Death, $1 \rightarrow 2'$ & 5FU+Lev & 0.235 & 0.113 &1.265& 0.037\\ 
& Extent 3 or 4 & 0.304 & 0.179 &1.355 & 0.091   \\
& Nodes $>4$ & 0.379 &  0.103 & 1.461 &$< 0.001$ \\
\midrule
Entry to Death, $0 \rightarrow 2$& 5FU+Lev & 0.031 & 0.333 &1.035 & 0.917\\ 
& Extent 3 or 4 & 0.108 & 0.449 & 1.115 &  0.809  \\
& Nodes $>4$ & 0.486 &  0.373 & 1.627 &  0.193 \\
\bottomrule
\end{tabular}
\end{table}

\section{Other Modeling Considerations\label{section2}}

\subsection{Delayed Entry and Incomplete Data on Process History}

The previous section covered intensity-based modeling in an idealized scenario, where processes were tracked from their onset. While this is typical in inception cohort studies, sometimes, individuals are enrolled in the study after the process has already been underway for some time.

Let $A_i$ denote the recruitment time of individual $i$, after which we plan to observe their process over the interval $(A_i, \tau]$.
In some settings,  such as the UK Biobank, data \citep{sudlow2015uk,gorfine2021marginalized} information on all transitions occurring over $(0, A_i]$ is available, in which case one can let $Y_i^A(t) = I(A_i \le t)$.
Then $\tilde{Y}_i(t) = Y_i^A(t) Y_i(t) Y_i^\dagger(t^-)$ indicates that individual $i$ is under study (i.e. has been recruited and has not yet been censored or entered an absorbing state), and
     $\tilde{Y}_{ik}(t) = Y_i^A(t) Y_i(t) I(Z_i(t^-)=k)$ indicates that the individual is under study and at risk for transition out of state $k$ at time $t$.
Similar to previous notation, let $d \tilde{N}_{ikl}(t) = \tilde{Y}_{ik}(t) dN_{ikl}(t)$, and the vectors
 $\tilde{N}_{ik}(t)=\left( \tilde{N}_{ikl}(t), l \neq k, l=1,\ldots,K\right)^{\sf T}$ and
 $ \tilde{N}_i(t)=\left(\tilde{N}^{\sf T}_{ik}(t),\notin \mathcal{A}\right)^{\sf T}$.
Then with time-fixed covariates, the broadened history, including the information on the delayed-entry time, may be written as
$\tilde{\cal H}_i(t) = \{Y_i^A(s), Y_i(s), d\tilde{N}_i(s), 0 \leq s < t, X_i\}$.

Under conditionally independent delayed entry  \citep{keiding1992} and conditionally independent loss to follow-up
\begin{equation*}
\lim_{\Delta t \downarrow 0} \displaystyle\frac{1}{\Delta t}{ \Pr \left(\Delta \tilde{N}_{ikl}(t)=1 | Z_i(t^-)=k, \tilde{\cal H}_i(t)\right)} = \tilde{Y}_{ik}(t) \lambda_{kl}\left(t|\tilde{\cal H}_i(t)\right) \, .
\label{delayed}
\end{equation*}
The likelihood can then be constructed similarly to Equations \eqref{likelihood}--\eqref{Lkl}, but with $\bar{Y}_{ik}(t)$ replaced by $\tilde{Y}_{ik}(t)$, $D_{ikl}$ being the set of transition times observed over $[A_i, \tau)$, and the integral starting at $A_i$ rather than 0.

In settings where information on the process for $t < A_i$ is either completely missing or  highly coarsened \citep{heitjan1993ignorability}  fitting intensity-based models that heavily depend  on the process history becomes challenging. While Markov models should be justified based on scientific  plausibility and evidence of adequacy, they are particularly appealing for use in such situations, as the intensities are independent of the histories given the current state.

\subsection{Time-Dependent Covariates and Joint Modeling}
The factorization of the likelihood given in Section \ref{sec1.4} justifies the use of the likelihood based solely on the multistate process. However, joint modeling of covariates and multistate processes is of scientific value in settings where the interest lies in the relationship between the two processes. For example, when studying the role of bone health markers in relation to the risk of fractures, continuous markers of bone formation and resorption could be incorporated into the intensities for the occurrence of first and subsequent bone fractures. Fractures, in turn, can affect bone markers, and this effect can be examined within a joint model for the two processes  \citep{cook2014statistical}. In this case, a likelihood based on  (\ref{eq-full-msm-LX}) and (\ref{eq-full-msm-LN}) can be considered. If the continuous bone markers are discretized, a joint multistate model can be constructed, with states defined by combinations of marker levels and fracture states, potentially including an absorbing state for death. In this example, an additional challenge arises when covariates are subject to intermittent observation. We discuss how this can be addressed in Section \ref{sec3}.

\subsection{Inference for Marginal Parameters and Pseudo-values}

The intensity functions are the fundamental components of a multistate process, and, as shown above, specifying all intensity functions enables the construction of the likelihood. This further implies that all probabilistic aspects of the process are determined — at least when the intensity model does not include time-dependent covariates that introduce `extra randomness' beyond the multistate process itself (i.e., when the likelihood based on Eq. (\ref{eq-full-msm-LX}) is straightforward).  Thus, marginal features, such as state occupancy probabilities, $p_k(t)$, and expected sojourn times, $\mu_k$,  in the states, can be estimated based on the estimated intensity functions, either through a `plug-in' approach (if the mathematical relationship can be specified) or via simulation.

However, in a regression setting, the plug-in approach does not provide parameters that directly describe the association between time-fixed covariates, $X$ and, for example, $p_k(t)$. Additionally, if the primary scientific interest lies in such an association, then typically \emph{all} intensity functions need to be modeled, and model misspecification becomes a concern. It is therefore of interest to directly specify a marginal model for the association, i.e., without relying on models for the intensity functions.

It may not be possible to specify intensity functions for the multistate process in such a way that a simple marginal model, such as (\ref{eq:GLMforp}) below, holds. Therefore, a direct marginal model should be seen as a `working model',  useful for assessing a direct association between the marginal parameter and $X$, but not necessarily reflecting the true data-generating mechanism. For any regression model, simplifying assumptions, such as additivity and linearity, should be carefully evaluated through appropriate diagnostics.

Here, we discuss two recent approaches to direct marginal modeling: \emph{direct binomial regression} using inverse probability of censoring weighted (IPCW) generalized estimating equations (GEE) \citep{directbinomial, ts-mjz-sjs07} and the \emph{pseudo-values} (PV) method \citep{pseudo-bka03, pka-maja-smmr10}; see also the recent book by \cite{PKAHRbook}. We illustrate these approaches by studying $p_k(t_0)$ for a fixed time point $t_0$, but emphasize that similar methods can be applied for joint inference at multiple time points,  $p_k(t_0),\dots,p_k(t_m)$, or for the $\tau-$restricted mean sojourn time  in state $k$, $\mu_k(\tau)=\int_0^{\tau}p_k(u)du$. Additionally, conditional probabilities, such as $p_{kl}(s,t|X) = \Pr(Z(t)=l \mid Z(s)=k, X)$ or $p_{kl}(s,t|X(s)) = \Pr(Z(t)=l \mid Z(s)=k, X(s))$ can be studied using these approaches via the method of \emph{landmarking} \citep{HansvHbook,putter2018non,LIDAOakes}.

Consider a regression model
\begin{equation}
g\left(p_k(t_0\mid X_i)\right)=X_i^{\sf T}\beta
\label{eq:GLMforp}
\end{equation}
where $p_k(t|X_i)=\Pr (Z_i(t)=k|Z_i(0)=0,X_i(0)=X_i)$, $g$ is a specified link function, and the coefficient vector $\beta$ includes an intercept specific to the time point $t_0$. Thus, the coefficients will be specific to both the state, $k$, and the time point, $t_0$, though for ease of notation we denote coefficient vector as $\beta$. Typical link functions include the cloglog, corresponding to a proportional hazards model in the two-state model (Fig. \ref{fig1} a)), or the logit function. 

Direct binomial regression builds on those subjects for whom the state $k$ indicator $I(Z_i(t_0)=k)$ at time $t_0$ is observed. These are the subjects with $t_0\wedge T_i^{\dagger}\leq C_i^{\ast}$, i.e., either $t_0$ or $T_i^{\dagger}=\inf_t \{Z_i(t)\in {\cal A}\}$  (the time at which $Z_i(t)$ reaches an absorbing state) must occur before the time $C_i^{\ast}$ of random censoring. The state indicators $I(Z_i(t_0)=k)$ for these subjects are then used as responses in a GEE, $U_D(\beta)=0$, with 
\begin{equation}
U_D(\beta)=\sum_{i=1}^n I(t_0\wedge T_i^{\dagger}\leq C_i^{\ast})W_i(t_0\wedge T_i^{\dagger})A(\beta,X_i)
\bigl\{I(Z_i(t_0)=k)-p_k(t_0\mid X_i)\bigr\} \, .
\label{eq:dirbinGEE}
\end{equation}
Each term has a  weight reflecting the probability $W_i(t)^{-1}=\Pr(C_i^{\ast}>t\mid X_i)$ of being uncensored, and typically $A(\beta,X_i)$ contains the partial derivatives of $p_k(t_0\mid X_i)=g^{-1}\left(\beta^{\sf T}X_i\right)$ with respect to $\beta$. Clearly, this approach requires a model for the random censoring $C^{\ast}$, and in its simplest form, the resulting weights could be given by the Kaplan-Meier estimator. However, if covariates affect censoring, a regression model would be needed for estimating the weights. The terms in (\ref{eq:dirbinGEE}) are independent, and a sandwich estimator for the variance of the solution to (\ref{eq:dirbinGEE}), denoted by $\widehat{\beta}_D$, is typically used, with a contribution arising from the need to estimate $W_i(\cdot)$ \citep{ts-mjz-sjs07}.

The marginal regression model (\ref{eq:GLMforp}) can also be analyzed using PV. With this approach, an outcome variable for each observation $i$, to be used in a GEE, is computed via a \emph{base estimator} of the marginal state occupancy probability $p_k(t_0)$, denoted by $\widehat{p}_k(t_0)$. The Aalen-Johansen estimator is consistent, even without assuming the multistate process is Markov \citep{Datta-Satten01, overgaard-19}.  The PV for subject $i$ is given by
\begin{equation}
V_i=n\widehat{p}_k(t_0)-(n-1)\widehat{p}_k^{(-i)}(t_0)
=\widehat{p}_k(t_0)+(n-1)\bigl\{\widehat{p}_k(t_0)-\widehat{p}_k^{(-i)}(t_0)\bigr\} \, ,
\label{eq:pseudo}
\end{equation}
where $\widehat{p}_k^{(-i)}$ is the (Aalen-Johansen) estimator applied to the sample of size $n-1$ obtained by removing subject $i$ from the full sample. In the special case of no censoring, the Aalen-Johansen estimator reduces to the relative frequency $\sum_{i=1}^n I(Z_i(t_0)=k)/n$ of processes in state $k$ at time $t_0$, and $V_i$ is then simply $I(Z_i(t_0)=k)$. Note that $V_i$ is calculated by (\ref{eq:pseudo}) for \emph{all} the subjects, even if $I(Z_i(t_0)=k)$ is observed.  The PV is then used as response in a GEE, $U_P(\beta)=0$, where
\begin{equation}
U_P(\beta)=\sum_{i=1}^n A(\beta,X_i)\bigl\{V_i-p_k(t_0\mid X_i)\bigr\} \, .
\label{eq:PVGEE}
\end{equation}
The terms in $U_P(\beta)$ are not independent \citep{OPP-pseudo}, and special techniques are needed for evaluating large-sample properties as $n\rightarrow\infty$ of the solution $\widehat{\beta}_P$ to (\ref{eq:PVGEE}). These depend on the properties of the influence function of the functional (denoted $\phi$) that maps the data from the observed multistate process onto $\widehat{p}_k(t_0)$ \citep{OPP-pseudo,OAP-stata}. A necessary condition for the properties to hold is that censoring does not depend on covariates. If covariates affect censoring, the Aalen-Johansen base estimator in (\ref{eq:pseudo}) may be replaced by an IPCW estimator of $p_k(t_0)$ \citep{OPP-covdepcens}. It should be noted that the required properties of the influence function are typically not fulfilled when the base estimator is based on data with delayed entry \citep{PAP-IJpseudo}. Another consequence of the lack of independence among the terms in (\ref{eq:PVGEE}) is that the standard GEE sandwich estimator for the variance of $\widehat{\beta}_P$ should be replaced by a corrected estimator that also involves the second-order influence function of the functional $\phi$. However, in practical applications, the correction terms tend to be small \citep{OPP-pseudo}.

No systematic comparison between the estimators $\widehat{\beta}_D$ and $\widehat{\beta}_P$, based on  (\ref{eq:dirbinGEE}) or  (\ref{eq:PVGEE}), has been conducted. In large samples, the computation of PV using (\ref{eq:pseudo}) can be time-consuming. Approximations via \emph{infinitesimal jackknife} PV method \citep{PAP-IJpseudo}, as implemented in the {\tt survival} package in {\tt R}, offer a more efficient alternative. Additionally, for certain specific multistate models, such as those shown in Fig. \ref{fig1} a), b), and f), direct models for all time points are available. These models each require specialized estimating equations, such as those based on partial likelihood principles \citep{fine-gray-99, lawless-nadeau,lin-etal-jrss-2000,ghosh-lin-02}.

\section{Intermittent Observation of Continuous-Time Processes}
\label{sec3}
In many settings, transitions between states are not directly observed, and only the state occupied at intermittent assessment times is recorded. Examples include studies of retinopathy where visual acuity is assessed during clinic visits \citep{marshall1993}, diabetic hepatology \citep{poynard1997}  where liver function is evaluated through blood tests or biopsies, and studies of osteoporosis where periodic radiographic examinations can detect asymptomatic vertebral fractures \citep{saag2018}. To accommodate intermittent observation in the likelihood construction, we consider the multistate process $\{Z_i(s),0 \leq s\}$ along with time-independent covariates $X_i$.  The assessment process is represented by a counting process $\{A_i(s),0 \leq s\}$, which records the number of assessments up to time $s$. An assessment at time $t$ results in $dA_i(t)=A_i(t)-A_i(t^-)=1$, while $dA_i(t)=0$ otherwise. Since visits can only occur for individuals still on the study, the assessment process is right-censored by $C_i$, so we observe $d\bar A_i(t)=Y_i(t) dA_i(t)$ and $\bar A_i(t)=\int_0^t d\bar A_i(s)$. If $\bar{\cal H}_i(t)=\{Y_i(s), \bar{A}_i(s), d \bar{N}_i(s), 0 \leq s < t, X_i\}$, then the assessment-process intensity is 
\begin{equation*}
\lim_{\Delta t\downarrow0}\frac{1}{\Delta t} {\Pr \left(\Delta A_i(t)=1|\bar{\cal H}_i(t) \right)}
\label{visit-intensity}
\end{equation*}
where $\Delta A_i(t)=A_i((t+\Delta t)^-) - A_i(t^-)$. This general formulation involves a dependence on $\{Z_i(s),0 \leq s < t\}$, but this process is not fully observed.  
In such cases, joint models for the disease and visit processes must be specified. These models are often constructed under the assumption of
conditional independence  given latent variables with an assumed distribution.  \cite{lange2015} and  \cite{cook2021} propose joint models that account for local dependence and, along with \cite{gruger1991} discuss the independence conditions needed to focus on the partial likelihood contributions involving only the intensities of the multistate process. Assume individual $i$ has $m_i$ visits at times $ 0 \leq a_{i0} < a_{i1} < \cdots < a_{im_i}$ and  let 
$\bar{\cal H}_i^\circ(t)=\{Y_i(s), \bar{A}_i(s), 0 \leq s < t, (Z_i(a_r), a_r), r=0, 1,\ldots, \bar{A}_i(t^{-}),X_i\}$ 
represent the observed  history of individual $i$ at time $t$. If the visit process is noninformative, meaning no  parameters are shared between the visit and multistate models, we can ignore  the visit process and  focus on a likelihood of the form
\begin{equation}\label{eq:grunger}
\prod_{r=1}^{m_i}\Pr \left(Z_i(a_r)|a_r,\bar {\cal H}_i^\circ(a_r)\right) \, .
\end{equation}
Expressing
$\Pr\left(Z_i(a_r)|a_r,\bar {\cal H}_i^\circ(a_r)\right)$ in terms of intensity functions can be challenging for general processes, but
Markov models  are relatively easy to handle. For instance,
if $\lambda_{kl}(t|X_i) = \lambda_{kl}(X_i) = \lambda_{kl}\exp(X_i^{\sf T}\beta_{kl})$ for all $k \rightarrow l $ transitions, we can construct a transition intensity matrix ${\mathbf Q}(X_i)$,  where $\lambda_{kl}(X_i)$ appears in the off-diagonals, and the diagonal elements are
$ - \sum_{l=0, l \neq k}^K \lambda_{kl}(X_i)$.
Then, the transition probability matrix
${\mathbf P}(s, t|X_i)=
\exp\{ {\mathbf Q}(X_i)(t-s)\}$ has entries $p_{kl}(s,t|X_i) = \Pr(Z_i(t)=l|Z_i(s)=k, X_i)$.
\cite{kalbfleisch1985} discuss a Fisher-scoring algorithm for such models, which, along with other optimization methods, 
is implemented in the  {\tt msm} \citep{msm} package in {\tt R}. The assumption of time-homogeneous baseline intensities can be relaxed allow them to be piecewise-constant rates upon specifying the number and location of cut points.

\cite{titman2010, titman2015} considers semi-Markov models with sojourn times having phase-type distributions; see  \cite{yang2011}.
\cite{satten1999} considers progressive models with Markov intensities given a common multiplicative random effect which accommodates both serial dependence in the sojourn times and heterogeneity across individuals in the progression rate. Extensions to accommodate more general forms of heterogeneity have also been developed to include higher dimensional random
effects \citep{sutradhar2008} and mover-stayer components \citep{okeeffe2013}.

Transition information is often available through dual observation schemes. For instance, in dementia studies that model cognition and survival, cognitive state is observed only during assessments, while survival times are recorded continuously. A simple model to illustrate this setup is the illness-death model shown in Fig. \ref{fig1}c).  If the assessment process for state 1 involves random visit times, the entry time to state 1 is interval-censored. Additionally, in some studies, there may be uncertainty about whether a transition has occurred. For example, if the disease had not been diagnosed by the last visit, it is unclear whether it developed between that visit and the time of death. Therefore, the likelihood must be adjusted to account for this uncertainty.
In this context,  \cite{leffondre2013} explore the usefulness of the illness-death model when the primary interest is overall survival. \cite{joly2002} develop methods for fitting intensity-based models to such data, using spline-based approaches for modeling the intensities or the Weibull distribution \citep{joly2012}.

The inclusion of time-dependent explanatory variables requires additional assumptions when these variables are not measured continuously. As a result, this situation is typically addressed using time-dependent but interval-constant covariates, where the times of changes in value are known.  \cite{boruvka2016} examine identifiability issues and apply a sieve maximum likelihood approach to estimate transition intensities and covariates' effects. More generally,  \cite{commenges2002} explore the estimation of multistate processes under intermittent observation using splines.

\subsection{The Psoriatic Arthritis Data Revisited}
\begin{figure}
\centering
\includegraphics[width=0.9\textwidth]{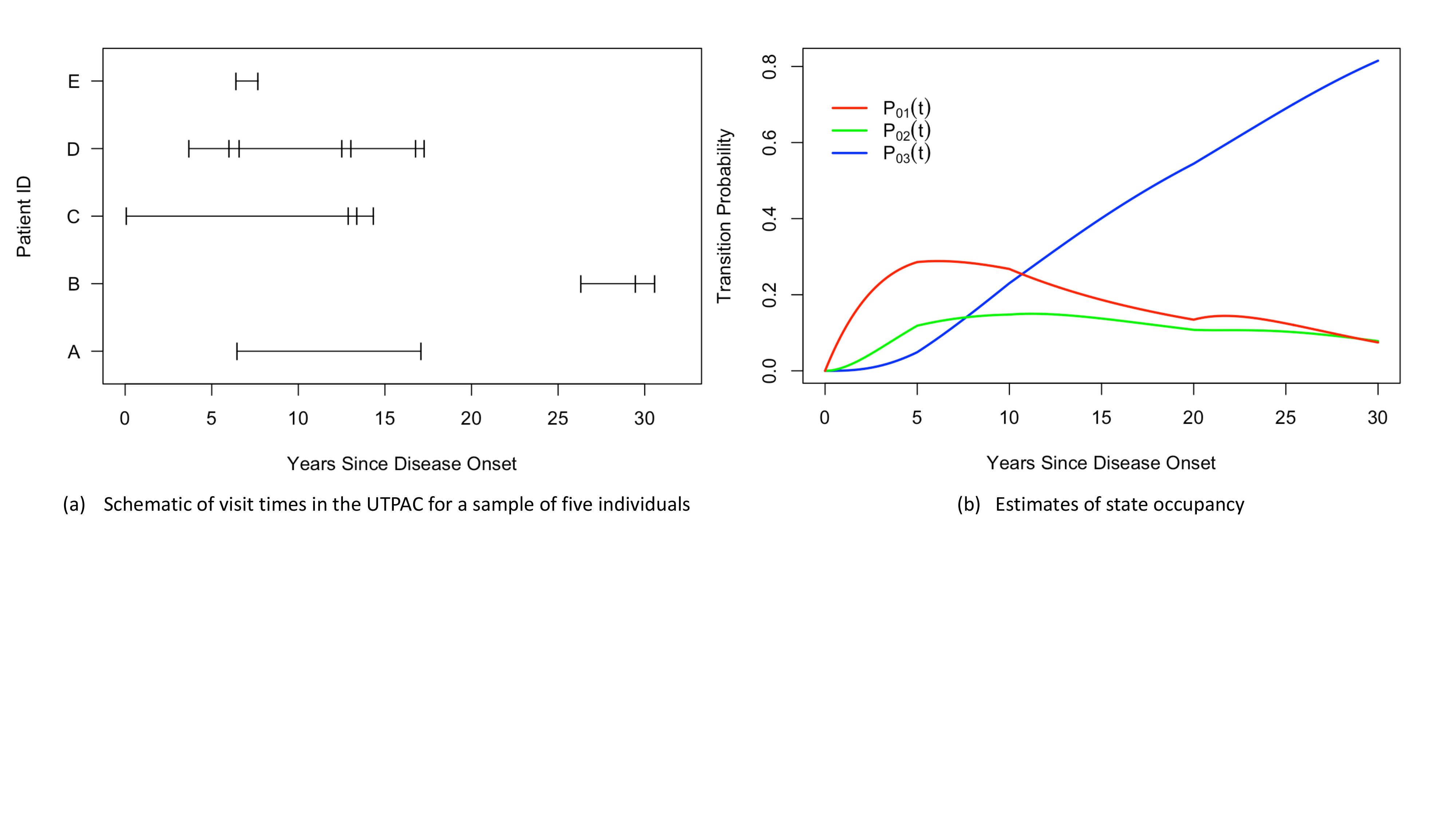}
\caption{Joint damage data from the University of Toronto Psoriatic Arthritis Cohort: sample data and estimates of state occupancy based on fitted Markov models with piecewise-constant baseline intensities having cut-points at 5, 10 and 20 years from the onset of psoriatic arthritis.}\label{fig:psoriatic}
\end{figure}

Here, we analyze the joint damage data in psoriatic arthritis from Section 1.2.2. Patients in this registry are scheduled for annual clinical examinations and biennial radiographic exams, but visit times vary greatly across and within patients. Fig. \ref{fig:psoriatic}(a) shows raw data from five patients, with horizontal lines representing the period from the first clinic visit to loss to follow-up or death. Vertical hatch marks indicate radiographic exams, highlighting the significant variability in the frequency of imaging data collection among individuals. We consider a four-state model from  \cite{gladman1995} as shown in Fig. \ref{fig1}b). Specifically, state 0 corresponds to having no damaged joints, state 1 to 1-4 damaged joints, state 2 to 5-9 damaged joints, and state 3 is an absorbing state representing 10 or more damaged joints. The time scale starts at the age of psoriatic arthritis diagnosis. Assuming a noninformative visit process, we use the likelihood (\ref{eq:grunger}) to estimate piecewise-constant transition intensities under Markov models, with cut-points at 5, 10, and 20 years after disease onset.  Fig. \ref{fig:psoriatic}(b) and Table \ref{tab:combined} are based on the  code provided in Section S3 of the Supplementary Material file.

Fig. \ref{fig:psoriatic}b) is obtained by maximizing the likelihood with respect to all intensity parameters $\lambda = (\lambda_{01}^{\sf T}, \lambda_{12}^{\sf T}, \lambda_{23}^{\sf T})^{\sf T}$ while omitting any covariates, where each $\lambda_{kl}$ is a $4 \times 1$ vector of parameters of the piecewise-constant intensity for a $k \rightarrow l$ transition with cut-points at 5, 10 and 20 years since disease diagnosis.
Using these estimates, we compute $P(0,t;\widehat{\lambda})$ and plot $p_k(t; \widehat{\lambda})$  over the 30 years following disease onset. The state occupancy probabilities for the transient states rise and then fall as patients progress to more severe joint damage, with nearly 50\% expected to reach 10 or more damaged joints after 20 years. Table \ref{tab:combined}  summarizes the results of intensity-based regression models, including baseline covariates - an indicator of extensive effusions and an indicator of an elevated erythrocyte sedimentation rate (ESR), a marker of inflammation. Table \ref{tab:combined} (a) shows the regression coefficient for each transition, and Table \ref{tab:combined} (b) provides the estimated transition intensities for each  time interval.
Interestingly, an elevated ESR at baseline is a strong predictor of a faster transition from $1 \rightarrow 2$, but it does not significantly impact the other two transitions.

\begin{table}[h]
\small
\centering
\caption{Estimates from Fitting Markov Models to data on joint damage from the University of Toronto Psoriatic Arthritis Cohort with piecewise-constant baseline intensities having cutpoints at 5, 10 and 20 years from the onset of  psoriatic arthritis }
\label{tab:combined}
\begin{subtable}[t]{1\textwidth}
\centering
\captionsetup{font=large}
 \caption{ Estimated regression coefficients (Est) and hazard rates (HR)}
\begin{tabular}{@{}lrrrrrr@{}} 
\toprule
Transition & Covariate & Est & SE & HR & 95\% CI & p-value \\
\midrule
$0 \rightarrow 1$  & Effusions & 0.742 & 0.400 & 2.100 & (0.960 , 4.597) & 0.063 \\
 & Elevated ESR & 0.239 & 0.278 & 1.271 & (0.737 , 2.188) & 0.389 \\
\midrule
$1 \rightarrow 2$  & Effusions & 0.536 & 0.297 & 1.710 & (0.955 , 3.062) & 0.071 \\
 & Elevated ESR & 0.774 & 0.281 & 2.169 & (1.250 , 3.759) & 0.006 \\
\midrule
$2 \rightarrow 3$ & Effusions & 0.306 & 0.311 & 1.358 & (0.739 , 2.497) & 0.325 \\
 & Elevated ESR & -0.359 & 0.364 & 0.698 & (0.342 , 1.425) & 0.323 \\
\bottomrule
\end{tabular}
\end{subtable}

\vspace{2mm} 

\begin{subtable}[t]{1\textwidth}
\centering
\captionsetup{font=large}
\caption{Estimated transition intensities (Intns), baselines are with covariates set to 0}
\begin{tabular}{ccccccccc} 
\toprule
 & \multicolumn{2}{c}{Baseline} & \multicolumn{2}{c}{$[5,10)$} &
  \multicolumn{2}{c}{$[10,20)$}  & \multicolumn{2}{c}{$[20,\infty)$}  \\
\midrule
 Transition  & Intns & 95\% CI & Intns & 95\% CI & Intns & 95\% CI & Intns & 95\% CI\\
\midrule
 $0 \rightarrow 1$  & 0.092 & (0.052 , 0.161) & 0.718 & (0.387 , 1.333) & 0.419 & (0.187 , 0.939) & 1.566 & (0.763 , 3.215)\\
 $1 \rightarrow 2$  & 0.097 & (0.048 , 0.198) & 0.828 & (0.394 , 1.743) & 0.796 &(0.406 , 1.559) & 1.126 & (0.502 , 2.529)\\
 $2 \rightarrow 3$  & 0.244 & (0.072 , 0.821) & 1.326  & (0.434 , 4.048) & 1.174 &(0.381 , 3.618) & 1.373 & (0.431 , 4.375)\\ 
\bottomrule
\end{tabular}
\end{subtable}
\end{table}

\section{Multistate Models with Frailty or Copula Approaches}

At times, even after accounting for covariates in order to explain between-individuals variation in disease course, life trajectories display greater variability than anticipated. To address this unobserved heterogeneity, models with latent random effects can be considered. In survival analysis, these are often referred to as frailty models \citep{vaupel1979impact}. Alternatively, copula models can be employed, where the joint multivariate distribution is constructed using the marginal distributions and a copula function \citep{shih1995inferences}. In general, these methods are commonly used to capture the influence of unobserved characteristics specific to individuals or clusters. Within the framework of multistate survival models, two key scenarios are relevant:
\begin{itemize}
\item[(i)]  \textit{Within-subject dependence:} Here, random effects or copula models account for unobserved covariates that influence event times within the same individual.  For instance, in Fig. \ref{fig1}d), a random effect might explain the unobserved dependence between the transition times of $0 \rightarrow 1$ and $1 \rightarrow 2'$.
\item[(ii)] \textit{Between-subjects dependence:} In this case, clustered data, such as families or study centers,  involve correlated failure times among individuals within the same cluster. Random effects or copula models can address this unobserved dependence.
\end{itemize}

Both scenarios will be discussed in the following sections. Applying frailty or copula models in the general multistate framework can be complex, as unobserved heterogeneity may vary across different transitions. We focus primarily on the illness-death model (Fig. \ref{fig1}c) or d)) for simplicity. As shown in Sections 5.1 and 5.2, within- and between-subject dependence issues remain largely unresolved in most multistate models.

\subsection{Within-Subject Random Effect}  
Consider the illness-death model for a chronic disease, which involves three possible transitions: health to disease ($0 \rightarrow 1$), health to disease-free death ($0 \rightarrow 2$), and disease to post-disease death  ($1 \rightarrow 2'$). The conditional independence assumption between time to disease diagnosis and time to death, given the observed covariates, is often unrealistic. This motivates the inclusion of an unobserved random subject-level effect in the model, which induces dependence among failure times of a subject. Conditional on the  observed covariates and unobserved random effect,  failure times within a subject are assumed to be independent.  

The following discussion examines two distinct approaches: one focuses on the regression coefficients of the observed covariates conditioned on the unobserved random effects, while the other considers the regression coefficients of the observed covariates without conditioning on the unobserved random effects. Each approach provides a different interpretation of the effect of the observed covariates, as detailed below. These approaches will be demonstrated using the Cox model (Section 5.1.1) and the accelerated failure time model (Section 5.1.2). Software implementation is outline in Section 6.7

Random-effects models for failure time outcomes are commonly referred to as frailty models, where the random subject-specific factor representing the unobserved risk associated with a subject’s failure times is called the frailty.  \cite{xu2010statistical} considered a one-parameter gamma-frailty model, in which a single frailty variate is shared across all three processes of the illness-death model.  While this approach has been widely adopted by subsequent researchers \citep{lee2015bayesian,haneuse2016semi,lee2017accelerated,jiang2017semi,gorfine2021marginalized,kats2023accelerated}, it could be viewed as imposing a strong assumption of identical unobserved dependence across the processes. In more complex multistate models with additional processes, a one-parameter frailty model becomes less realistic. An alternative is to use a vector-valued frailty variate for each subject \citep{liquet2012investigating,cook-lawless-book2018} or introduce a transformation for the frailty effect corresponding to each process \citep{liquet2012investigating}. To our knowledge, most existing frailty approaches for multistate models focus on illness-death and progressive (forward) multistate models, as shown in Fig. \ref{fig1}b), including extensions with competing risks and two-level clustering \citep{jung2019joint}.

A key feature of the frailty approach is the assumption that, given the observed covariates and the frailty variate, individual transition processes are independent. This assumption allows for the separate modeling of each transition process. Also, it is often assumed that the unobserved frailty variate and the observed covariates are independent.
However, certain assumptions are required to ensure model identifiability \citep{putter2015frailties} and verifying these assumptions with available data can be challenging. When addressing unobservable random effects, it is important to distinguish between two approaches: conditional modeling, which conditions on both observed covariates and the unobserved frailty variate, and marginal (population-average) modeling, which conditions only on the observed covariates. In linear models, if the observed covariates are independent of the frailty variate, these approaches yield the same estimates. However, in nonlinear models, they do not, making the distinction practically significant. The choice between the two approaches depends on the specific objectives of the analysis.

The choice of time scale in multistate models with frailty is crucial. Consider two illness-death models: in the first, the states are ``Healthy,'' (state 0) ``Disease,'' (1) and ``Death'' (2 and $2'$). In the second, the states are ``Surgery,'' ``Recurrence,'' and ``Death.'' In the first case, a negative association is expected between age at diagnosis and the time from diagnosis to death, as older individuals generally have shorter lifespans post-diagnosis. Therefore,  given our intention to utilize a shared random effect to capture the interplay between the three processes, it
is natural to utilize the age scale for all three processes. In the second case, a clock-reset approach is needed, where time resets at each transition, in case early recurrence is  positively associated with the remaining lifespan. Here, the time scale for each transition depends on the time spent in the preceding state.

\subsubsection{Within-Subject Random Effects: Conditional vs Marginalized Illness-Death Cox Models}
Since the illness-death model of Fig. \ref{fig1} d) describes a progressive process where each state is visited at most once, we adopt the simplified notation mentioned in Section 1. Let there be $n$ independent observations, with $T_{i1}$ and $T_{i2}$ denoting the times to disease diagnosis and death, respectively, $i=1,\ldots,n$. Let $C_i$ represent the right-censoring time, and $X_{i}$ a set of time-fixed covariates.  While some of the models discussed here can be easily extended to incorporate time-dependent covariates, others would require substantial additional modifications. Define $W_{i1}=\min(T_{i1},T_{i2},C_i)$ as the earliest occurrence time among disease diagnosis, death, or censoring. Let $\delta_{i1}=I(W_{i1}=T_{i1})$ indicate whether $W_{i1}$ corresponds to disease diagnosis, and let $\delta_{i2}=I(W_{i1}=T_{i2})$ whether it corresponds to death. Additionally, let $W_{i2}=\delta_{i1} \min(T_{i2},C_i)$ represent the time to death or censoring, given that the disease diagnosis is observed, or 0 otherwise. 
Let  $\delta_{i3}$ indicate whether the terminal event occurred with $\delta_{i3}=\delta_{i1}I(W_{i2}=T_{i2})$. The observed data is then $\{W_{i1},W_{i2},\delta_{i1},\delta_{i2},\delta_{i3},X_{i} , ; , i=1,\ldots,n\}$. Denote the unobserved frailty variate of individual $i$ by a random variable $\omega_i > 0$ with a cumulative distribution $F_{\omega}$ indexed by an unknown parameter $\theta$.

\cite{xu2010statistical} proposed an illness-death model featuring three Cox-based hazard functions. One of their key innovations was the inclusion of a gamma-distributed shared frailty variate, which applies a multiplicative effect to each intensity function. This accommodates unobserved dependencies between the times to nonterminal and terminal events. The conditional intensity functions governing the three transition processes are expressed as
\begin{eqnarray}\label{eq:coxcond1}
    \lambda_{01}(t|X_{i},\omega_i) &=& \lim_{\Delta t \rightarrow 0} (\Delta t)^{-1} \Pr(\Delta N_{i01}(t)=1|Z_i(t^-)=0,X_i,w_i) \nonumber \\
    &=&\lim_{\Delta t \rightarrow 0} (\Delta t)^{-1} \Pr (t \leq T_{i1} < t+\Delta t|T_{i1} \geq t, T_{i2} \geq t, X_{i},\omega_i) \nonumber \\
    &=&
\omega_i \lambda_{01}(t) \exp\left(X_i^{\sf T}\beta_{01} \right) \,\,\,\,\,\,   t >0
\end{eqnarray}
\begin{eqnarray}\label{eq:coxcond2}
\lambda_{02}(t|X_{i},\omega_i) &=& \lim_{\Delta t \rightarrow 0} (\Delta t)^{-1} \Pr(\Delta N_{i02}(t)=1|Z_i(t^-)=0,X_i,w_i)
\nonumber \\
 &=& \lim_{\Delta t\rightarrow 0} (\Delta t)^{-1} \Pr (t \leq T_{i2} < t+\Delta t|T_{i1} \geq t, T_{i2} \geq t, X_{i},\omega_i) \nonumber \\
 &=&
\omega_i \lambda_{02}(t) \exp\left(X_i^{\sf T}\beta_{02} \right)    \,\,\,\,\,\,   t >0
\end{eqnarray}
\begin{eqnarray}\label{eq:coxcond3}
\lambda_{12'}(s|t,X_{i},\omega_i) &=& \lim_{\Delta s\rightarrow 0} (\Delta s)^{-1} 
\Pr(\Delta N_{i12}(s)=1|T_{i1} = t, Z_i(s^-)=1,X_i,w_i)
\nonumber \\
&=& \lim_{\Delta s\rightarrow 0} (\Delta s)^{-1} \Pr (s \leq T_{i2} < s+\Delta s|T_{i1} = t, T_{i2} \geq s, X_{i},\omega_i) \nonumber \\
&=&
\omega_i \lambda_{12'}(s) \exp\left( X_i^{\sf T}\beta_{12'}\right)   \,\,\,\,\,\,  s > t >0  
\end{eqnarray}
where $\lambda_{kl}$ and $\beta_{kl}$, $kl \in \{01,02,12'\}$, are transition-specific baseline hazard functions and vectors of regression coefficients, respectively. Given that subject $i$ was diagnosed at time $T_{i1}=t$, the support of $T_{i2}$ is restricted by $t$, so the conditional distribution of $T_{i2}$ is truncated by $t$. The time  to the nonterminal event, $T_{i1}=t$, is not incorporated into the covariate vector for $\lambda_{12'}$. Rather, the connection between the potential event times $T_{i1}$ and $T_{i2}$ is derived from two main factors: the so-called explanatory hazard ratio $\lambda_{12'}/\lambda_{02}$  \citep{xu2010statistical}, and the frailty variate $\omega_i$. Should $T_{i1}$ and $T_{i2}$ be independent, it is expected that the explanatory hazard ratio remains constant at 1 throughout, and similarly, the frailty variate should also maintain a value of 1.

Estimation of models (\ref{eq:coxcond1})--(\ref{eq:coxcond3}) under gamma-distributed frailty with mean 1 and unknown variance can be done  by semiparametric maximum-likelihood  estimators (MLE) \citep{xu2010statistical} or by a semiparametric Bayesian approach \citep{lee2015bayesian}. Interesting modifications of (\ref{eq:coxcond1})-(\ref{eq:coxcond3})  can be found in  \cite{jiang2017semi} and \cite{lee2021fitting}.

Survival predictions based on (\ref{eq:coxcond1})--(\ref{eq:coxcond3}) require knowledge of both the observed covariates and the unobserved frailty variate, which limits their practical applicability. By integrating out the frailty, we obtain the so-called {\it marginalized hazards} with respect to $\omega_i$. These marginalized hazards are heavily influenced by the choice of frailty distribution and depend on its parameters. This approach is often less desirable, as it obscures the interpretation of covariate effects and increases sensitivity to misspecifications in the frailty distribution.
To address these issues, \cite{gorfine2021marginalized} proposed an alternative strategy for frailty-based illness-death models, where the marginal hazards are modeled using Cox models, and the conditional hazards are derived for a given frailty distribution. These conditional hazards, which depend on $(X_i,\omega_i)$ and the parameters of the marginal hazards, depart from the proportional-hazards structure. This approach allows for the estimation of marginal model parameters while incorporating frailties to account for unobserved subject-specific covariates. Specifically, the conditional hazards  are given by 
\begin{equation}\label{eq:cond1}
\lambda_{0l}(t|X_{i},\omega_i) = 
\omega_i \alpha_{01}(t| X_i) \hspace{0.3cm} l=1,2 \hspace{0.3cm} t>0
\end{equation}
\begin{equation}\label{eq:cond2}
\lambda_{12'}(s|t,X_{i},\omega_i) = 
\omega_i \alpha_{12'}(s| X_i) \hspace{0.3cm} s>t>0 \, .    
\end{equation}
The corresponding marginalized hazard functions are defined to be
\begin{equation}\label{eq:marg1}
\lambda_{0l}(t|X_{i}) = \lambda_{0l}(t) \exp\left(X_i^{\sf T}\beta_{0l} \right) \hspace{0.3cm} l=1,2 \hspace{0.3cm} t>0    
\end{equation}
\begin{equation}\label{eq:marg2}
\lambda_{12'}(s|t,X_{i}) = \lambda_{12'}(s) \exp\left(X_i^{\sf T}\beta_{12'} \right) \hspace{0.3cm} s>t>0    
\end{equation}
with unspecified baseline hazard functions $\lambda_{kl}(\cdot)$, $kl \in \{01,02,12'\}$. The nonnegative functions $\alpha_{kl}(\cdot|X_i)$  are determined by the distribution of the frailty $\omega_i$ and the corresponding marginalized hazard $\lambda_{kl}(\cdot|X_i)$. For example, under the gamma-frailty model with expectation 1 and variance $\theta$, 
$\alpha_{kl}(t|X_i) = \lambda_{kl}(t) \alpha^*_{kl}(t|X_i) \hspace{0.3cm}$,  $kl \in \{01,02,12\}$
where
$\alpha^{*}_{0l}(t | X_i) = \exp \left(X_i^{\mathsf{T}} \beta_{0l}\right) \exp\left\{\theta \Lambda_{0.}(t |X_i)\right\}$, $l = 1,2$, 
$\Lambda_{0.}(t |X_i)=\Lambda_{01}(t |X_i)+\Lambda_{02}(t |X_i)$ and
$\alpha^{*}_{12'}(t|X_i) = \exp\left(X_i^{\sf T}
\beta_{12'}\right)\exp \left\{\Lambda_{12'}(t|X_i)\theta/(1+\theta)\right\}/(1+\theta)$.
In summary, the above frailty-based models (\ref{eq:cond1}) and (\ref{eq:cond2}) that depart from a proportional hazards structure are expressed in terms of the main parameters of interest, $\beta_{kl}$ and $\lambda_{kl}(\cdot)$ of models (\ref{eq:marg1}) and (\ref{eq:marg2}). The frailty-based models are particularly appealing for developing estimation procedures, as they assume that, given the observed covariates and the frailty variate, individual transition processes are conditionally independent. A pseudo-likelihood approach was developed for estimating the parameters under standard frailty-specific assumptions \citep{gorfine2021marginalized}.

\subsubsection{Within-Subject Random Effect: Additive vs Multiplicative Illness-Death AFT Models}
An important alternative of the Cox proportional hazards model is the accelerated failure time (AFT) regression. Instead of modeling the intensity function, the AFT model models the logarithm  of survival times as a linear function of covariates. In the context of illness-death models with frailty, \cite{lee2017accelerated} proposed
the following {\it additive scale-change} model
\begin{eqnarray*}
	\log(T_{i1})&=&X_i^{\sf T}\beta_{01} +\omega_i+U_{i01} \, ,\, T_{i1}>0\\
	\log(T_{i2})&=&X_i^{\sf T}\beta_{02} +\omega_i+U_{i02} \, ,\, T_{i2}>0\, , \,\mbox{given subject $i$ is free of the disease}\\
	\log(T_{i2})&=&X_i^{\sf T}\beta_{12'} +\omega_i+U_{i12'} \, ,\, T_{i2}>t_{1}>0, \,\mbox{given subject $i$ was diagnosed at time $T_{i1}=t_1$}
\end{eqnarray*}
where additivity is with respect to the shared-frailty component $\omega_i$.  The random errors $U_{ikl}$ are independent across $i=1,\ldots,n$ and $kl \in \{01,02,12'\}$, possibly having unspecified distributions.  The association between $T_{i1}$ and $T_{i2}$ is determined by the distribution of $\omega_i$, manifesting as an additive component in the log-failure times scale. Given the assumption of $\omega_i$ following a normal distribution, both parametric and semiparametric Bayesian estimation methods are available \citep{lee2017accelerated}.

Alternatively, a {\it multiplicative} frailty-based AFT model has been proposed \cite{kats2023accelerated},  where the unobserved frailty variate is not directly expressed in the log-failure time linear model, but instead affects the distribution of the random errors $U_{ikl}$. In particular, let
\begin{eqnarray*}
	\log(T_{i1})&=&X_i^{\sf T} \beta_{01} +U_{i01} \, ,  \, T_{i1}>0\\
	\log(T_{i2})&=&X_{i}^{\sf T} \beta_{02} +U_{i02} \, ,  \, T_{i2}>0 \, ,\,\mbox{given subject $i$ is free of the disease}\\
	\log(T_{i2})&=&X_{i}^{\sf T}\beta_{12'} +U_{i12'} \, ,  \, T_{i2}>t_{i1}>0 \, ,\,\mbox{given subject $i$ was diagnosed at time $T_{i1}=t_1$} \, .
\end{eqnarray*}
The dependence between $T_{i1}$ and $T_{i2}$ is incorporated via the following shared frailty model. Given individual $i$'s frailty variate $\omega_i$, it is assumed that the respective conditional baseline hazard functions of $\exp(U_{ikl})$, $kl \in \{01,02,12'\}$, are given by
\begin{eqnarray*}
	\lambda_{01}(t|\omega_i) &=& \omega_i \lambda_{01}(t) \, , \, t>0    \\
	\lambda_{02}(t|\omega_i) &=& \omega_i \lambda_{02}(t) \, , \, t>0 \, ,  \mbox{given subject $i$ is free of the disease} \\
	\lambda_{12'}(s|t_1,\omega_i) &=& \omega_i \lambda_{12'}(s) \, , \, s>t_1>0 \, , \mbox{given subject $i$ was diagnosed at time $T_{i1}=t_1$}
\end{eqnarray*}
where each $\lambda_{kl}(\cdot)$ is an unspecified baseline hazard function of $\exp(U_{il})$. Under the assumption that $\omega_i$ are gamma distributed with mean 1 and unknown variance $\theta$, a semiparametric MLE (based on a kernel-smoothed likelihood combined with an EM algorithm) is available \cite{kats2023accelerated}. Section 2.2 of \cite{kats2023accelerated} delves into the conceptual distinctions between the above additive and multiplicative approaches. It is elucidated that the hazards of the additive approach may exhibit non-monotonic behavior with respect to $\omega_i$. In contrast, the hazards of the multiplicative model demonstrate monotonic increase as a function of  $\omega_i$ across all error distributions. Consequently, the multiplicative-frailty model offers a simpler interpretation.

\subsubsection{The Rotterdam Tumor Bank Data Revisited}
Table \ref{tab7} provides a summary of the analysis of the Rotterdam tumor bank data, of Section 1.2.3, based on the codes of Section S4 of the Supplementary Material file. The analysis includes three different frailty-based models: the multiplicative AFT model \citep{kats2023accelerated}, the marginalized Cox model \cite{gorfine2021marginalized}, and the conditional Cox model \citep{lee2015bayesian}. The AFT additive-frailty model \citep{lee2017accelerated} implemented in the {\texttt R} package \texttt{SemicompRisks}, uses the sojourn time $T_{i2}-T_{i1}$ when death occurs after relapse. This model encountered convergence issues when applied to the data. We believe this is due to the potential negative correlation between sojourn time and the time from surgery to relapse, while the gamma frailty model assumes a positive correlation. The multiplicative AFT model suggests a strong dependence between the time to relapse and time to death, not explained by the observed covariates. Similarly, the marginalized Cox model indicates a high level of dependence between these events, while the conditional Cox model shows a somewhat weaker relationship. Although the direction of the covariates' effects is consistent across the Cox models and the AFT model, the inference results vary slightly between the three models.  A visual goodness-of-fit assessment demonstrates that models without frailty show a poorer fit to the data compared to those incorporating frailty \citep{kats2023accelerated}. It should be noted that time to relapse, $T_{i1}=t$, is not included in the covariate vector for transition $1 \rightarrow 2'$ in these frailty approaches. As a result, the association between the time to relapse and time to death is captures solely through the frailty variate, as explained earlier. This simplifies the interpretation of the results, as the frailty variate provides a unified framework for understanding the dependence between these events given the observed covariates. 

\begin{sidewaystable}
\spacingset{1.2}
	\caption{Rotterdam Tumor Bank Data: Estimates (Est) or posterior medians (PM), standard errors (SE), RR, and credible intervals at credibility level $0.05$ for the hazard-ratio parameters under the Bayesian approach. }\label{tab7}
	\centering
	\scalebox{0.75}{
		\begin{tabular}{lccccccccc}
			& \multicolumn{3}{c}{AFT-Multipliation \citep{kats2023accelerated}}        
			& \multicolumn{3}{c}{Cox-Marginalized \citep{gorfine2021marginalized}}        
			& \multicolumn{3}{c}{Cox-Conditional \citep{lee2015bayesian}} \\
			& Est (SE) & RR & p-value         
			& Est (SE) & RR & p-value        
			& PM (SE) & RR & Credible Interval \\
			\midrule
			$\theta$ & 2.18 (0.73) &    -  & 0.003        
			         & 2.52 (0.54) &    -  & 0.000         
			         & 1.47 (0.23) & - & (1.046,1.956) \\
			\midrule
			\textbf{Transition: surgery $\rightarrow$ relapse, $0 \rightarrow 1$} \\
			Age at surgery (divided by 10) &  0.14 (0.06)  & 1.15  & 0.012             
			                               &  -0.15 (0.06) & 0.86  & 0.014  
			                               &  -0.22 (0.08) & 0.80 &  {(0.685,0.918)} \\
			log of lymph nodes & -0.40 (0.05) & 0.67  & 0.000 
			                   & 0.42 (0.04)  & 1.53  & 0.000   
			                   & 0.71 (0.07)  & 2.03 & {(1.795,2.326)}\\
			log of estrogen+1  & 0.07 (0.03)  & 1.07  & 0.030  
			                   & -0.03 (0.02) & 0.97  & 0.186   
			                   & -0.10 (0.04) & 0.90  & {(0.839,0.964)}  \\
			log of progesterone+1 & 0.09 (0.02)  & 1.09  & 0.000 
			                      & -0.04 (0.02) & 0.96  & 0.065 
			                      & -0.11 (0.03) & 0.90  & {(0.845,0.958)} \\
			Postmenopausal (vs. premenopausal) & -0.34 (0.15) & 0.71  & 0.023 
			                                   & 0.13 (0.13)  & 1.14  & 0.296   
			                                   & 0.34 (0.19)  & 1.40  & (0.980,2.081) \\
			Tumor size (ref $<20$mm) \\
			\,\,\,\,\, 20-50mm   & -0.32 (0.09)  & 0.73  & 0.001 
		                                       &  0.20 (0.07)  & 1.22  & 0.006     
		                                       &  0.40 (0.12)  & 1.49  & {(1.180,1.882)} \\
		     \,\,\,\,\, $>50$mm   & -0.49 (0.11) & 0.61  & 0.000 
	                                           &  0.38 (0.11)  & 1.46  & 0.001 
	                                           &  0.79 (0.16)  & 2.19  & {(1.625,3.007)}\\
			Hormone therapy &  0.60 (0.13)  & 1.83  & 0.000 
			                & -0.38 (0.08) & 0.68  & 0.000 
			                &  -0.88 (0.15) & 0.41 & {(0.310,0.541)} \\
			Chemotherapy &   0.49 (0.11) & 1.64   & 0.000 
			             &  -0.37 (0.11) & 0.69  & 0.001 
			             &  -0.79 (0.16) & 0.46  & {(0.329,0.615)}\\
			Tumor grade 3 (vs. 2) & -0.25 (0.09) & 0.78  & 0.004 
			                      & 0.21 (0.08) & 1.23  & 0.008 
			                      & 0.44 (0.13) & 1.56 & {(1.216,1.986)}\\
			\midrule
			\textbf{Transition: surgery $\rightarrow$ death, $0 \rightarrow 2$} \\
			Age at surgery (divided by 10) & -0.43 (0.14) & 0.65  & 0.002 
			                               & 1.32 (0.37) & 3.74  & 0.000 
			                               & 1.43 (0.18) & 4.20  & {(2.987,5.923)} \\
			log of lymph nodes & -0.14 (0.08) & 0.87  & 0.091 
			                   & 0.13 (0.12) & 1.14  & 0.298 
			                   & 0.44 (0.15) & 1.54  & {(1.163,2.092)} \\
			log of estrogen+1  & 0.04 (0.04) & 1.04 & 0.287 
			                   & -0.01 (0.06) & 0.99 & 0.816 
			                   & -0.11 (0.08) & 0.89  & (0.765,1.040) \\
			log of progesterone+1 & 0.01 (0.04)  & 1.01  & 0.827 
			                      & 0.08 (0.06) & 1.08  & 0.205 
			                      & 0.01 (0.07) & 1.01  & (0.884,1.163) \\
			Postmenopausal (vs. premenopausal)	& -0.15 (0.34) & 0.86  & 0.647                             
												& -0.30 (0.50) & 0.74  & 0.554     
												& -0.35 (0.70) & 0.70  & (0.179,2.997)         \\
			Tumor size (ref. $<20$mm)\\
			\,\,\,\,\, 20--50mm 	& -0.13 (0.15) & 0.88  & 0.376                      
												& -0.16 (0.25) & 0.85  & 0.526      
												& -0.04 (0.28) & 0.96 & (0.554,1.653) \\
			\,\,\,\,\, $>50$mm  & -0.19 (0.18) & 0.82  & 0.275                 
											& 0.15 (0.31)  & 1.16  & 0.634 
											& 0.58 (0.35) & 1.79  & (0.933,3.488) \\
			Hormone therapy 	& 0.41 (0.18) & 1.51 & 0.019                   
							& -0.21 (0.25) & 0.81  & 0.389 
							& -0.69 (0.29) & 0.50  & {(0.275,0.851)} \\
			Chemotherapy    	& 1.13 (0.30) & 3.09  & 0.000                   
							& -0.22 (0.81) & 0.81  & 0.789 
							& -0.78 (0.63) & 0.46  & (0.130,1.531) \\
		    Tumor grade 3 (vs. 2) 	& -0.06 (0.13) & 0.94  & 0.641         
									& -0.01 (0.28) & 0.99  & 0.961        
									& 0.21 (0.27) & 1.23  & (0.750,2.148)\\
			\midrule
		    \textbf{Transition: relapse $\rightarrow$ death, $1 \rightarrow 2'$}  \\
			Age at surgery (divided by 10)	& 0.00 (0.07) & 1.00  & 0.956         
											& 0.03 (0.08) & 1.03 & 0.700        
											& 0.08 (0.07) & 1.08  & (0.931,1.232)          \\
			log of lymph nodes 	& -0.25 (0.07) & 0.78  & 0.000 
								& 0.25 (0.05) & 1.28 & 0.000 
								& 0.38 (0.07)  & 1.47  & {(1.271,1.687)} \\
			log of estrogen+1  	& 0.04 (0.05) & 1.04 & 0.341          
								& -0.03 (0.02) & 0.97  & 0.193          
								& -0.10 (0.04) & 0.90  & {(0.838,0.973)}          \\
			log of progesterone+1 	& 0.13 (0.04)  & 1.14  & 0.001          
									& -0.08 (0.02) & 0.92  & 0.000 
									& -0.19 (0.04) & 0.83  & {(0.771,0.884)} \\
			Postmenopausal (vs. premenopausal)  	& -0.21 (0.17) & 0.81  & 0.203        
												& -0.05 (0.13) & 0.95  & 0.731       
												& 0.04 (0.20) & 1.04  & (0.705,1.527)        \\
		    Tumor size (ref. $<20$mm)  \\
		    \,\,\,\,\, 20--50mm 	& -0.37 (0.14) & 0.69  & 0.008          
												& 0.23 (0.07)  & 1.26  & 0.001 
												& 0.46 (0.14) & 1.58  & {(1.234,2.112)} \\
			\,\,\,\,\,  $>50$mm  & -0.52 (0.17) & 0.60 & 0.002          
											& 0.40 (0.10)  & 1.49 & 0.000 
											& 0.67 (0.18) & 1.96  & {(1.405,2.764)} \\
			Hormone therapy	& 0.39 (0.14) & 1.48  & 0.005         
							& -0.18 (0.09) & 0.84  & 0.037         
							& -0.48 (0.16) & 0.62 & {(0.452,0.835)} \\
			Chemotherapy   	& 0.23 (0.18) & 1.25  & 0.205         
							& -0.16 (0.13) & 0.85  & 0.227        
							& -0.18 (0.17) & 0.84   & (0.604,1.179)         \\
			Tumor grade 3 (vs. 2) 	& -0.26 (0.13) & 0.77  & 0.047      
									& 0.21 (0.09) & 1.23  & 0.024     
									& 0.43 (0.14) & 1.54 & {(1.177,2.034)}\\
			\midrule
	\end{tabular}}
\end{sidewaystable}

\subsection{Between-Subjects Dependence}

There has been limited exploration of illness-death models in the context of clustered data, such as family or twin studies.  Frailty-based models and estimation methods for competing risks (i.e., Fig. \ref{fig1} f) in clustered failure-time data have been developed \citep{joly2012,bandeen2002modelling,gorfine2011frailty,cederkvist2019modeling}. However, extending these approaches to more complex settings of multistate models is challenging, primarily due to the varying strength of dependence between two cluster members across different transitions. \cite{lee2021fitting} developed an illness-death model using the latent variable formulation of the competing-risk model for the first transition, $0 \rightarrow 1$ or $0 \rightarrow 2$, and a copula model is adopted to accommodate dependence within clusters in the (possibly latent)  times to transition from state 0 to 1.

\section{Software Availability}
This section offers an overview of the software and packages (in either R or Python) accessible for analyzing multistate models, highlighting the distinctive contributions of each package.

\subsection{R Packages \texttt{survival} and \texttt{mstate}}
The \texttt{survfit} function \citep{therneau2024multi} of the package \texttt{survival} \citep{therneau2013modeling,therneau2020package} calculates and plots the Aalen-Johansen estimators of cumulative incident functions within any multistate scenario. In particular, subjects can visit multiple states during the course of a study, subjects can start after time 0 (i.e., delayed entry), and
they can start in any of the states. The standard error of the Aalen-Johansen estimates is computed using an infinitesimal jackknife. The \texttt{coxph} function offers Cox regression analysis for each transition in a multistate model, with or without shared coefficients. The \texttt{mstate} package \citep{de2010mstate,de2011mstate} incorporates utilities for data preparation, descriptive analyses, hazard functions estimation,  prediction using Aalen-Johansen estimator and Cox regression modeling. Due to its modular approach, different models, like for instance additive hazards models, can be fitted for the transition intensities, while still allowing prediction based on Aalen-Johansen. Also, functions for testing the Markov assumption  are provided \citep{titman2020general}.

\subsection{R package \texttt{msm}}
The \texttt{msm} package \citep{msm}  is designed for analyzing continuous-time Markov models with piecewise-constant intensity functions, as well as hidden Markov models that include unobserved (hidden) states; the latter are not discussed here. The package provides estimates for transition intensities, transition probability matrices, and expected time spent in each state. It also yields predictions for future state occupancy. Parameters are estimated via maximum likelihood, enabling standard inferential techniques for confidence intervals and hypothesis testing.

\subsection{Python Package \texttt{PyMSM}}
\texttt{PyMSM} \citep{rossman2022pymsm} is a package for fitting competing risks and multistate models, offering flexible model specification, individual and population-level predictions, and comprehensive statistical summaries and visualizations. Key features include: (1) Multistate Regression Model Fitting: Supports a variety of survival analysis techniques, such as Cox regression, random survival forests \citep{ishwaran2014random}, or user-defined machine learning models.
(2) Prediction via Monte Carlo Simulation: Using a fitted multistate model, \texttt{PyMSM} generates sample paths through Monte Carlo simulations. Given covariates, initial states, and times, it sequentially samples subsequent states and durations in each state based on the estimated model, ending when a terminal state is reached or a predefined maximum number of transitions is exceeded. Summary statistics, such as state occupancy probabilities and median state durations, are available after sampling multiple paths per observation.
(3) Predefined Models and Data Simulation: Allows loading or configuring predefined multistate models and generating simulated survival data via random paths, providing a valuable tool for research.

\subsection{R Package \texttt{SmoothHazard}}
The \texttt{SmoothHazard} package \citep{touraine2017smoothhazard} is designed for fitting regression models to interval-censored data within illness-death models. It includes algorithms for concurrently fitting regression models to the three transition intensities of an illness-death model, where the transition times to State 1 (see Fig. \ref{fig1}c) may be interval-censored, and all event times can be right-censored. The three baseline transition intensity functions are modeled either by Weibull distributions or, alternatively, by M-splines in a semi-parametric framework. Given specific covariates, the estimated transition intensities can be combined to produce estimates of cumulative incidence functions and life expectancies.

\subsection{R packages \texttt{pseudo} and \texttt{eventglm}}
The \texttt{pseudo} R package \citep{pseudo} includes functions for computing pseudo-values (see Section 2.3) for various marginal estimands of interest such as the cumulative incidence
function, or the restricted mean time in a state. The R package \texttt{eventglm} also applies the pseudo-values framework and includes plotting of
residuals, the use of sampling weights, and corrected
variance estimation.

\subsection{R package \texttt{simMSM}}
The R package \texttt{simMSM} \citep{simMSM} simulates event histories for multistate models. It enables the generation of event histories featuring potentially non-linear baseline hazard functions, as well as nonlinear time-dependent or time-independent covariates' effect, while also accounting for dependencies on past history. The random generation of event histories is achieved through inversion sampling applied to cumulative all-cause hazard rate functions.

\subsection{R Packages Targeting Illness-Death Models with Within-Subject Random Effects}
The R package \texttt{SemiCompRisks} \citep{semiCompRisks} employs Bayesian estimation techniques for frailty-based illness-death models, encompassing both conditional and additive models as delineated in Sections 5.1.1-5.1.2. This package offers Cox-type and accelerated failure time  models incorporating gamma and normal frailty distributions, respectively.  The \texttt{frailty-LTRC}  package (available at \url{https://github.com/nirkeret/frailty-LTRC})  utilizes a pseudo-likelihood approach for marginalized Cox models outlined in Section 5.1.1, employing a gamma frailty. On the other hand, \texttt{semicompAFT} (available at \url{https://github.com/leakats/semicompaft}) implements a semi-parametric AFT model under the multiplicative frailty setting discussed in Section 5.1.2, also employing a gamma frailty distribution. \texttt{frailtypack} \citep{frailtypack} is a package focusing particularly on handling frailty models and is versatile for time-to-event data in complex scenarios, including multistate models with recurrent events or competing risks.

\section{Discussion}
In this paper, we have presented various approaches to modeling multistate processes, with a focus on both intensity-based and marginal models. Sections 2 and 3  play a crucial role in laying the foundational concepts for understanding these models and their practical utility. Section 4 builds on this by addressing the challenges posed by intermittent observation in continuous-time processes, which is a frequent issue in clinical studies where transitions between  states are not continuously observed. Our exploration of frailty-based models in Section 5 illustrates their ability to capture unobserved heterogeneity. The approaches discussed in this paper provide critical tools for handling the complexities of real-world data, enabling researchers to make more accurate inferences about the dynamics of processes such as disease progression.

While the  frailty-based models provide a useful framework for addressing subject-specific unobserved covariates, they also introduce additional complexities, especially regarding the assumptions about dependence structures. Future research should focus on extending these methods to accommodate more complex multistate processes. Additionally, further development of software tools that can handle these advanced models will be essential for making these techniques  accessible to researchers in various fields.

Another important area not covered here is the application of machine learning methods to multistate survival data. Although the use of machine learning in multistate survival analysis is still emerging \citep{lee2018deephit,ishwaran2007random,bou2011review}, these approaches show considerable potential for improving prediction accuracy, particularly in healthcare. The main challenge is to balance the predictive power of machine learning with the need for interpretability and robustness, which are crucial for clinical decision-making. As the field advances, the integration of machine learning into multistate modeling frameworks will likely open new avenues for analyzing complex survival data. 

Validating multistate models presents significant challenges, particularly when evaluating their predictive performance or assessing goodness-of-fit with real-world data. A comprehensive literature review of existing model checking methods, along with an identification of the key unresolved issues, would be highly valuable for advancing this area.

\section*{Supplementary Material}

\section*{Acknowledgments}
The work of M.G. was supported by the Israel Science Foundation, grant number 767/21, and by
a grant from the Tel Aviv University Center for AI and Data Science (TAD).
R.J.C. was supported by a Discovery Grant from the Natural Sciences and Engineering Research Council of Canada (RGPIN-2017-04207) and the Canadian Institutes of Health Research (FRN 13887). M.A. is a Distinguished James McGill Professor of Biostatistics at McGill University.

\section*{Appendix - Glossary of Notation}
$\mathcal{S}$ - the set of states.\\
$\mathcal{A}$ - the set of absorbing states.\\
$Z(t)$ - the state occupied at time $t$, where $t$ is the time since the origin of a process.\\
$\{Z(s), 0 \leq s\}$ - the multistate process. \\
$H(t)=\{Z(s), 0 \leq s < t \}$ - the history of the process up to (not including) time $t$.\\
$T_k$ -  the entry time to state $k$, useful for progressive processes.\\
$T_k^{(r)}$ - the $r$th entry time to state $k$.\\
$N_{kl}(t)$ - the number of transitions $k \rightarrow l$ during $[0,t]$.\\
$\lambda_{kl}(t|H(t))$ - the $k \rightarrow l$ intensity function given the history $H(t)$.\\
$\Lambda_{kl}(t|H(t))$ - the cumulative transition intensity.\\
$p_{kl}(s,t)=\Pr(Z(t)=l|Z(s)=k)$ - the transition probability of $k \rightarrow l$.\\
$p_k(t)=\Pr(Z(t)=k|Z(0)=0)$ - the probability that state $k$ is occupied at time $t$.\\

\noindent
$C_i^*$ - a random right-censoring time of individual $i$.\\
$\tau$ - a fixed administrative censoring time.\\
$C_i = \min(C_i^*,\tau)$  \\
$Y_i(t) = I(t \leq C_i)$ - equals 1 if individual $i$ is uncensored prior to time $t$ and 0 otherwise.\\ 
$Y_{ik}(t) = I(Z_i(t)=k)$ - equals 1 if state $k$ is occupied at time $t$ by individual $i$.\\
$\bar{Y}_{ik}(t) = Y_i(t) Y_{ik}(t^-)$ - equals 1 if individual $i$ may be observed to transit out of state $k$ at time $t$.\\ 
$\bar{Y}_{\cdot k}(t) =\sum_{i=1}^n \bar{Y}_{ik}(t)$  - the total number of individuals at risk of a $k \rightarrow l$ transition at time $t$.\\
$N_{ikl}(t)$ - the total number of $k \rightarrow l$ transitions over $[0,t]$ of individual $i$.\\
$\Delta N_{ikl}(t) = N_{ikl}((t+\Delta t)^-)-N_{ikl}(t^-)$ - the number of events of individual $i$ over $[t, t+\Delta t)$.\\
$dN_{ikl}(t) = \lim_{\Delta t \downarrow 0} \Delta N_{ikl}(t)$ \\
$d \bar{N}_{ikl}(t) = \bar{Y}_{ik}(t) \, d N_{ikl}(t)$ - equals 1 if a $k \rightarrow l$ transition is recorded for process $i$ at time $t$.\\ $\bar{N}_{ikl}(t) = \int_0^t \bar{Y}_{ik}(s) \, d N_{ikl}(s)$ - the total number of observed $k \rightarrow l$ transitions of individual $i$  over the interval $[0, t]$. \\ 
$d\bar{N}_{\cdot kl}(t) = \sum_{i=1}^n \bar{Y}_{ik}(t) dN_{ikl}(t)$ - the total number of observed $k \rightarrow l$ transitions at time $t$.\\

\noindent
$X_i(t)$ - a $p\times 1$ vector of covariates at time $t$ of individual $i$.\\
$\{X_i(t), 0 \leq t\}$ - the covariates' path up to time $t$ of individual $i$.\\
$\mathcal{H}_i(t) = \{Z_i(s), X_i(s) , 0\leq s < t \}$ - the extended version of $H_i(t)$, includes the history of the multistate process and the covariates' path up to time $t$.\\

\noindent
$C_i^*(t)=I(C_i^* \leq t)$ - equals 1 if random censoring occurred by time $t$, and 0 otherwise.\\
$Y_i^{\dagger}(t)=I(Z_i(t) \notin \mathcal{A})$ - the occupied state at time $t$ by individual $i$ is a non-absorbing state.\\
$\bar{Y}_i(t)=Y_i(t)Y_i^{\dagger}(t^-)$ - equals 1 if individual $i$ may be observed to transit out of any state $k$ at time $t$, $k \notin \mathcal{A}$.\\
$\bar{N}_{ik}(t)=(\bar{N}_{ikl},l\neq k,l=1,\dots,K)^{\sf T}$ - vector of the cumulative number of transitions from state $k$ over $(0,t]$.\\
$\bar{N}_i(t)=(\bar{N}^{\sf T}_{ik}(t),k \notin \mathcal{A})^{\sf T}$ - vector of all counting processes of non-absorbing states.\\
$\Delta \bar{X}_i(t)=\bar{Y}_i(t+\Delta t) \{ X_i((t+\Delta t)^-) - X_i(t^-)\}$ - the increment in the covariate vector over $[t,t+\Delta t)$.\\
$d \bar{X}_i(t)=\lim_{\Delta t \downarrow 0} \Delta X_{i}(t)$\\
$\bar{X}_i(t)=\int_0^t d\bar{X}_i(s)$\\
$\bar{\mathcal{H}}_i(t) = \{Y_i(s), \bar{N}_i(s), \bar{X}_i(s) , 0\leq s < t; Z_i(0),X_i(0) \}$\\
$0=u_0 < u_1 < \cdots < u_R=\tau$ - a partition of $[0,\tau]$.\\
$\Delta \bar{X}_i(u_r) = \bar{Y}_i(u_r) ( X_i(u_r) - X_i(u_{r-1}) )  $ - an increment in the covariate vector of individual $i$. \\
$\Delta \bar{N}_i(u_r) Y_i(u_r) (N_i(u^-_r) - N_i(u^-_{r-1}) )$ - t the number of transitions of individual $i$ over the $[u_{r-1},u_r)$. \\
$\bar{{H}}_i(u_r) = \{Y_i(u_s), \Delta \bar{N}_i(u_s), \Delta \bar{X}_i(u_s) , s=1,\ldots,r; Z_i(0),X_i(0) \}$\\

\noindent
$A_i$ - the recruitment time of individual $i$.\\
$Y_i^{A}(t)=I(A_i \leq t)$ - individual $i$ is recruited before time $t$.\\
$\tilde{Y}_i(t)=Y_i^A(t)Y_i(t)Y_i^{\dagger}(t^-)$ - equals 1 if individual $i$ may be observed to transit out of any state $k$ at time $t$, $k \notin \mathcal{A}$.\\
$\tilde{Y}_{ik}(t)=Y_i^A(t)Y_i(t)I(Z_i(t^{-}=k)$ - the individual is under study and at risk of a transition out of state $k$ at time $t$.\\
$d \tilde{N}_{ikl}(t) = \tilde{Y}_{ik}(t)dN_{ikl}(t)$\\
$\tilde{N}_{ik}(t)=(\tilde{N}_{ikl},l\neq k,l=1,\dots,K)^{\sf T}$ - vector of the cumulative number of transitions from state $k$ over $(0,t]$.\\
$\tilde{N}_i(t)=(\tilde{N}^{\sf T}_{ik}(t),k \notin \mathcal{A})^{\sf T}$ - vector of all counting processes of non-absorbing states.\\
$\tilde{\mathcal{H}}_i(t) = \{Y^A_i(s), Y_i(s), d\tilde{N}_i(s), 0\leq s < t, X_i \}$\\

\noindent
$A_i(s)$ - the number of assessments of individual $i$ up to time $s$.\\
$dA_i(t) = A_i(t)-A_i(t^-)$ - equals 1 if an assessment occurred at time $t$, and 0 otherwise.\\
$d\bar{A}_i(t)=Y_{i}(t)A_i(t)$\\
$\bar{A}_i(t) = \int_0^t d \bar{A}_i(s)$\\
$\bar{\mathcal{H}}_i(t) = \{Y_i(s),\bar{A}_i(s), d\bar{N}_i(s), 0\leq s < t, X_i \}$\\
$0 \leq a_{i0} < a_{i1} < \cdots < a_{im_i}$ - assessments times of individual $i$.\\
$\bar{\cal H}_i^\circ(t)=\{Y_i(s), \bar{A}_i(s), 0 \leq s < t, (Z_i(a_r), a_r), r=0, 1,\ldots, \bar{A}_i(t^{-}),X_i\}$

\bibliography{literature}

\begin{thebibliography}{}

\bibitem[\protect\citeauthoryear{Aalen, Borgan, and Gjessing}{Aalen
  et~al.}{2008}]{aalen2008survival}
Aalen, O., {\O}.~Borgan, and H.~Gjessing (2008).
\newblock {\em Survival and Event History Analysis: a Process Point of View}.
\newblock Springer Science \& Business Media.

\bibitem[\protect\citeauthoryear{Aalen, Borgan, and Fekj{\ae}r}{Aalen
  et~al.}{2001}]{aalen2001covariate}
Aalen, O.~O., {\O}.~Borgan, and H.~Fekj{\ae}r (2001).
\newblock Covariate adjustment of event histories estimated from markov chains:
  the additive approach.

\bibitem[\protect\citeauthoryear{Aalen and Johansen}{Aalen and
  Johansen}{1978}]{aalen1978}
Aalen, O.~O. and S.~Johansen (1978).
\newblock An empirical transition matrix for non-homogeneous markov chains
  based on censored observations.
\newblock {\em Scandinavian Journal of Statistics\/}, 141--150.

\bibitem[\protect\citeauthoryear{Andersen, Borgan, Gill, and Keiding}{Andersen
  et~al.}{1993}]{andersen1993statistical}
Andersen, P.~K., {\O}.~Borgan, R.~D. Gill, and N.~Keiding (1993).
\newblock {\em Statistical Models Based on Counting Processes}.
\newblock Springer-Verlage New York, Inc.

\bibitem[\protect\citeauthoryear{Andersen, Klein, and Rosth{\o}j}{Andersen
  et~al.}{2003}]{pseudo-bka03}
Andersen, P.~K., J.~P. Klein, and S.~Rosth{\o}j (2003).
\newblock {Generalized linear models for correlated pseudo-observations, with
  applications to multi-state models}.
\newblock {\em Biometrika\/}~{\em 90}, 15--27.

\bibitem[\protect\citeauthoryear{Andersen and {Pohar Perme}}{Andersen and
  {Pohar Perme}}{2010}]{pka-maja-smmr10}
Andersen, P.~K. and M.~{Pohar Perme} (2010).
\newblock {{P}seudo-observations in survival analysis}.
\newblock {\em Statist. Meth. Med. Res.\/}~{\em 19}, 71--99.

\bibitem[\protect\citeauthoryear{Andersen and Ravn}{Andersen and
  Ravn}{2023}]{PKAHRbook}
Andersen, P.~K. and H.~Ravn (2023).
\newblock {\em Models for Multi-State Survival Data. Rates, Risks, and
  Pseudo-Values}.
\newblock Boca Raton: Chapman and Hall/CRC.

\bibitem[\protect\citeauthoryear{Andersen, Wandall, and {Pohar Perme}}{Andersen
  et~al.}{2022}]{LIDAOakes}
Andersen, P.~K., E.~N.~S. Wandall, and M.~{Pohar Perme} (2022).
\newblock {Inference for transition probabilities in non-Markov multi-state
  models}.
\newblock {\em Lifetime Data Analysis\/}~{\em 28}, 585--604.

\bibitem[\protect\citeauthoryear{Bandeen-Roche and Liang}{Bandeen-Roche and
  Liang}{2002}]{bandeen2002modelling}
Bandeen-Roche, K. and K.-Y. Liang (2002).
\newblock Modelling multivariate failure time associations in the presence of a
  competing risk.
\newblock {\em Biometrika\/}~{\em 89\/}(2), 299--314.

\bibitem[\protect\citeauthoryear{Beyersmann, Allignol, and
  Schumacher}{Beyersmann et~al.}{2011}]{beyersmann2011competing}
Beyersmann, J., A.~Allignol, and M.~Schumacher (2011).
\newblock {\em Competing Risks and Multistate Models with R}.
\newblock Springer Science \& Business Media.

\bibitem[\protect\citeauthoryear{Boruvka and Cook}{Boruvka and
  Cook}{2016}]{boruvka2016}
Boruvka, A. and R.~J. Cook (2016).
\newblock Sieve estimation in a markov illness-death process under dual
  censoring.
\newblock {\em Biostatistics\/}~{\em 17\/}(2), 350--363.

\bibitem[\protect\citeauthoryear{Bou-Hamad, Larocque, and Ben-Ameur}{Bou-Hamad
  et~al.}{2011}]{bou2011review}
Bou-Hamad, I., D.~Larocque, and H.~Ben-Ameur (2011).
\newblock A review of survival trees.
\newblock {\em Statistics Surveys\/}~{\em 5}, 44--71.

\bibitem[\protect\citeauthoryear{Cederkvist, Holst, Andersen, and
  Scheike}{Cederkvist et~al.}{2019}]{cederkvist2019modeling}
Cederkvist, L., K.~K. Holst, K.~K. Andersen, and T.~H. Scheike (2019).
\newblock Modeling the cumulative incidence function of multivariate competing
  risks data allowing for within-cluster dependence of risk and timing.
\newblock {\em Biostatistics\/}~{\em 20\/}(2), 199--217.

\bibitem[\protect\citeauthoryear{{Christopher H. Jackson}}{{Christopher H.
  Jackson}}{2011}]{msm}
{Christopher H. Jackson} (2011).
\newblock Multi-state models for panel data: The {msm} package for {R}.
\newblock {\em Journal of Statistical Software\/}~{\em 38\/}(8), 1--29.

\bibitem[\protect\citeauthoryear{Commenges}{Commenges}{2002}]{commenges2002}
Commenges, D. (2002).
\newblock Inference for multi-state models from interval-censored data.
\newblock {\em Statistical Methods in Medical Research\/}~{\em 11\/}(2),
  167--182.

\bibitem[\protect\citeauthoryear{Cook and Lawless}{Cook and
  Lawless}{2018}]{cook-lawless-book2018}
Cook, R. and J.~Lawless (2018).
\newblock {\em Multistate Models for the Analysis of Life History Data}.
\newblock Boca Raton, FL: CRC Press.

\bibitem[\protect\citeauthoryear{Cook}{Cook}{2014}]{cook2014statistical}
Cook, R.~J. (2014).
\newblock Statistical models for disease processes: Markers and skeletal
  complications in cancer metastatic to bone.
\newblock {\em Statistics in Action: A Canadian Outlook\/}, 177.

\bibitem[\protect\citeauthoryear{Cook and Lawless}{Cook and
  Lawless}{2007}]{cook2007}
Cook, R.~J. and J.~F. Lawless (2007).
\newblock {\em The Statistical Analysis of Recurrent Events}.
\newblock Springer.

\bibitem[\protect\citeauthoryear{Cook and Lawless}{Cook and
  Lawless}{2021}]{cook2021}
Cook, R.~J. and J.~F. Lawless (2021).
\newblock Independence conditions and the analysis of life history studies with
  intermittent observation.
\newblock {\em Biostatistics\/}~{\em 22\/}(3), 455--481.

\bibitem[\protect\citeauthoryear{Dabrowska}{Dabrowska}{1995}]{dabrowska1995estimation}
Dabrowska, D. (1995).
\newblock Estimation of transition probabilities and bootstrap in a
  semiparametric markov renewal model.
\newblock {\em Journal of Nonparametric Statistics\/}~{\em 5\/}(3), 237--259.

\bibitem[\protect\citeauthoryear{Datta and Satten}{Datta and
  Satten}{2001}]{Datta-Satten01}
Datta, S. and G.~A. Satten (2001).
\newblock Validity of the aalen-johansen estimators of stage occupation
  probabilities and nelson-aalen estimators of integrated transition hazards
  for non-markov models.
\newblock {\em Stat. \& Prob. Letters\/}~{\em 55}, 403--411.

\bibitem[\protect\citeauthoryear{De~Wreede, Fiocco, and Putter}{De~Wreede
  et~al.}{2010}]{de2010mstate}
De~Wreede, L.~C., M.~Fiocco, and H.~Putter (2010).
\newblock The mstate package for estimation and prediction in non-and
  semi-parametric multi-state and competing risks models.
\newblock {\em Computer Methods and Programs in Biomedicine\/}~{\em 99\/}(3),
  261--274.

\bibitem[\protect\citeauthoryear{de~Wreede, Fiocco, and Putter}{de~Wreede
  et~al.}{2011}]{de2011mstate}
de~Wreede, L.~C., M.~Fiocco, and H.~Putter (2011).
\newblock mstate: an r package for the analysis of competing risks and
  multi-state models.
\newblock {\em Journal of Statistical Software\/}~{\em 38}, 1--30.

\bibitem[\protect\citeauthoryear{Espenshade and Braun}{Espenshade and
  Braun}{1982}]{espenshade1982}
Espenshade, T.~J. and R.~E. Braun (1982).
\newblock Life course analysis and multistate demography: An application to
  marriage, divorce, and remarriage.
\newblock {\em Journal of Marriage and the Family\/}, 1025--1036.

\bibitem[\protect\citeauthoryear{Fine and Gray}{Fine and
  Gray}{1999}]{fine-gray-99}
Fine, J.~P. and R.~J. Gray (1999).
\newblock {A proportional hazards model for the subdistribution of a competing
  risk}.
\newblock {\em J. Amer. Statist. Assoc.\/}~{\em 94}, 496--509.

\bibitem[\protect\citeauthoryear{Gelber, Gelman, and Goldhirsch}{Gelber
  et~al.}{1990}]{gelber1990}
Gelber, R., R.~Gelman, and A.~Goldhirsch (1990).
\newblock A quality-of-life-oriented endpoint for comparing therapies.
\newblock {\em Maturitas\/}~{\em 12\/}(2), 152.

\bibitem[\protect\citeauthoryear{Ghosh and Lin}{Ghosh and
  Lin}{2002}]{ghosh-lin-02}
Ghosh, D. and D.~Y. Lin (2002).
\newblock {Marginal regression models for recurrent and terminal events}.
\newblock {\em Statistica Sinica\/}~{\em 12}, 663--688.

\bibitem[\protect\citeauthoryear{Gladman and Farewell}{Gladman and
  Farewell}{1995}]{gladman1995}
Gladman, D.~D. and V.~T. Farewell (1995).
\newblock The role of hla antigens as indicators of disease progression in
  psoriatic arthritis.
\newblock {\em Arthritis \& Rheumatism: Official Journal of the American
  College of Rheumatology\/}~{\em 38\/}(6), 845--850.

\bibitem[\protect\citeauthoryear{Gorfine and Hsu}{Gorfine and
  Hsu}{2011}]{gorfine2011frailty}
Gorfine, M. and L.~Hsu (2011).
\newblock Frailty-based competing risks model for multivariate survival data.
\newblock {\em Biometrics\/}~{\em 67\/}(2), 415--426.

\bibitem[\protect\citeauthoryear{Gorfine, Keret, Ben~Arie, Zucker, and
  Hsu}{Gorfine et~al.}{2021}]{gorfine2021marginalized}
Gorfine, M., N.~Keret, A.~Ben~Arie, D.~Zucker, and L.~Hsu (2021).
\newblock Marginalized frailty-based illness-death model: application to the
  uk-biobank survival data.
\newblock {\em Journal of the American Statistical Association\/}~{\em
  116\/}(535), 1155--1167.

\bibitem[\protect\citeauthoryear{Grossman, Mukherjee, Vaughan, Eastwood, Cook,
  LaForge, and Lampron}{Grossman et~al.}{1998}]{grossman1998}
Grossman, R., J.~Mukherjee, D.~Vaughan, C.~Eastwood, R.~Cook, J.~LaForge, and
  N.~Lampron (1998).
\newblock A 1-year community-based health economic study of ciprofloxacin vs
  usual antibiotic treatment in acute exacerbations of chronic bronchitis: the
  canadian ciprofloxacin health economic study group.
\newblock {\em Chest\/}~{\em 113\/}(1), 131--141.

\bibitem[\protect\citeauthoryear{Gr{\"u}ger, Kay, and Schumacher}{Gr{\"u}ger
  et~al.}{1991}]{gruger1991}
Gr{\"u}ger, J., R.~Kay, and M.~Schumacher (1991).
\newblock The validity of inferences based on incomplete observations in
  disease state models.
\newblock {\em Biometrics\/}, 595--605.

\bibitem[\protect\citeauthoryear{Haneuse and Lee}{Haneuse and
  Lee}{2016}]{haneuse2016semi}
Haneuse, S. and K.~H. Lee (2016).
\newblock Semi-competing risks data analysis: accounting for death as a
  competing risk when the outcome of interest is nonterminal.
\newblock {\em Circulation: Cardiovascular Quality and Outcomes\/}~{\em
  9\/}(3), 322--331.

\bibitem[\protect\citeauthoryear{Heitjan}{Heitjan}{1993}]{heitjan1993ignorability}
Heitjan, D.~F. (1993).
\newblock Ignorability and coarse data: Some biomedical examples.
\newblock {\em Biometrics\/}, 1099--1109.

\bibitem[\protect\citeauthoryear{Ishwaran, Gerds, Kogalur, Moore, Gange, and
  Lau}{Ishwaran et~al.}{2014}]{ishwaran2014random}
Ishwaran, H., T.~A. Gerds, U.~B. Kogalur, R.~D. Moore, S.~J. Gange, and B.~M.
  Lau (2014).
\newblock Random survival forests for competing risks.
\newblock {\em Biostatistics\/}~{\em 15\/}(4), 757--773.

\bibitem[\protect\citeauthoryear{Ishwaran and Kogalur}{Ishwaran and
  Kogalur}{2007}]{ishwaran2007random}
Ishwaran, H. and U.~B. Kogalur (2007).
\newblock Random survival forests for r.
\newblock {\em R News\/}~{\em 7\/}(2), 25--31.

\bibitem[\protect\citeauthoryear{Jiang and Haneuse}{Jiang and
  Haneuse}{2017}]{jiang2017semi}
Jiang, F. and S.~Haneuse (2017).
\newblock A semi-parametric transformation frailty model for semi-competing
  risks survival data.
\newblock {\em Scandinavian Journal of Statistics\/}~{\em 44\/}(1), 112--129.

\bibitem[\protect\citeauthoryear{Joly, Commenges, Helmer, and Letenneur}{Joly
  et~al.}{2002}]{joly2002}
Joly, P., D.~Commenges, C.~Helmer, and L.~Letenneur (2002).
\newblock A penalized likelihood approach for an illness--death model with
  interval-censored data: application to age-specific incidence of dementia.
\newblock {\em Biostatistics\/}~{\em 3\/}(3), 433--443.

\bibitem[\protect\citeauthoryear{Joly, Gerds, Qvist, Commenges, and
  Keiding}{Joly et~al.}{2012}]{joly2012}
Joly, P., T.~Gerds, V.~Qvist, D.~Commenges, and N.~Keiding (2012).
\newblock Estimating survival of dental fillings on the basis of
  interval‐censored data and multi‐state models.
\newblock {\em Statistics in Medicine\/}~{\em 31\/}(11-12), 1139--1149.

\bibitem[\protect\citeauthoryear{Jung, Peduzzi, Allore, Kyriakides, and
  Esserman}{Jung et~al.}{2019}]{jung2019joint}
Jung, T.~H., P.~Peduzzi, H.~Allore, T.~C. Kyriakides, and D.~Esserman (2019).
\newblock A joint model for recurrent events and a semi-competing risk in the
  presence of multi-level clustering.
\newblock {\em Statistical Methods in Medical Research\/}~{\em 28\/}(10-11),
  2897--2911.

\bibitem[\protect\citeauthoryear{Kalbfleisch and Lawless}{Kalbfleisch and
  Lawless}{1985}]{kalbfleisch1985}
Kalbfleisch, J. and J.~F. Lawless (1985).
\newblock The analysis of panel data under a markov assumption.
\newblock {\em Journal of the American Statistical Association\/}~{\em
  80\/}(392), 863--871.

\bibitem[\protect\citeauthoryear{Kats and Gorfine}{Kats and
  Gorfine}{2023}]{kats2023accelerated}
Kats, L. and M.~Gorfine (2023).
\newblock An accelerated failure time regression model for illness--death data:
  A frailty approach.
\newblock {\em Biometrics\/}~{\em 79\/}(4), 3066--3081.

\bibitem[\protect\citeauthoryear{Keiding and Moeschberger}{Keiding and
  Moeschberger}{1992}]{keiding1992}
Keiding, N. and M.~Moeschberger (1992).
\newblock {\em Independent Delayed Entry. In: Klein, John P. and Goel, Prem K.
  (eds) Survival Analysis: State of the Art}, pp.\  309--326.
\newblock Dordrecht: Springer Netherlands.

\bibitem[\protect\citeauthoryear{Kragh~Andersen, Pohar~Perme, van Houwelingen,
  Cook, Joly, Martinussen, Taylor, Abrahamowicz, and Therneau}{Kragh~Andersen
  et~al.}{2021}]{andersen2021}
Kragh~Andersen, P., M.~Pohar~Perme, H.~C. van Houwelingen, R.~J. Cook, P.~Joly,
  T.~Martinussen, J.~M. Taylor, M.~Abrahamowicz, and T.~M. Therneau (2021).
\newblock Analysis of time-to-event for observational studies: Guidance to the
  use of intensity models.
\newblock {\em Statistics in Medicine\/}~{\em 40\/}(1), 185--211.

\bibitem[\protect\citeauthoryear{Lange, Hubbard, Inoue, and Minin}{Lange
  et~al.}{2015}]{lange2015}
Lange, J.~M., R.~A. Hubbard, L.~Y. Inoue, and V.~N. Minin (2015).
\newblock A joint model for multistate disease processes and random informative
  observation times, with applications to electronic medical records data.
\newblock {\em Biometrics\/}~{\em 71\/}(1), 90--101.

\bibitem[\protect\citeauthoryear{Lawless and Cook}{Lawless and
  Cook}{2019}]{lawless2019new}
Lawless, J.~F. and R.~J. Cook (2019).
\newblock A new perspective on loss to follow-up in failure time and life
  history studies.
\newblock {\em Statistics in Medicine\/}~{\em 38\/}(23), 4583--4610.

\bibitem[\protect\citeauthoryear{Lawless and Nadeau}{Lawless and
  Nadeau}{1995}]{lawless-nadeau}
Lawless, J.~F. and J.~C. Nadeau (1995).
\newblock {Some simple robust methods for the analysis of recurrent events}.
\newblock {\em Technometrics\/}~{\em 37}, 158--168.

\bibitem[\protect\citeauthoryear{Lee, Gilsanz, and Haneuse}{Lee
  et~al.}{2021}]{lee2021fitting}
Lee, C., P.~Gilsanz, and S.~Haneuse (2021).
\newblock Fitting a shared frailty illness-death model to left-truncated
  semi-competing risks data to examine the impact of education level on
  incident dementia.
\newblock {\em BMC Medical Research Methodology\/}~{\em 21\/}(1), 1--13.

\bibitem[\protect\citeauthoryear{Lee, Zame, Yoon, and van~der Schaar}{Lee
  et~al.}{2018}]{lee2018deephit}
Lee, C., W.~R. Zame, J.~Yoon, and M.~van~der Schaar (2018).
\newblock Deephit: A deep learning approach to survival analysis with competing
  risks.
\newblock {\em Proceedings of the AAAI Conference on Artificial
  Intelligence\/}~{\em 32\/}(1).

\bibitem[\protect\citeauthoryear{Lee, Haneuse, Schrag, and Dominici}{Lee
  et~al.}{2015}]{lee2015bayesian}
Lee, K.~H., S.~Haneuse, D.~Schrag, and F.~Dominici (2015).
\newblock Bayesian semi-parametric analysis of semi-competing risks data:
  Investigating hospital readmission after a pancreatic cancer diagnosis.
\newblock {\em Journal of the Royal Statistical Society. Series C, Applied
  statistics\/}~{\em 64\/}(2), 253.

\bibitem[\protect\citeauthoryear{Lee, Lee, Alvares, and Haneuse}{Lee
  et~al.}{2021}]{semiCompRisks}
Lee, K.~H., C.~Lee, D.~Alvares, and S.~Haneuse (\, 2021).
\newblock {\em SemiCompRisks: Hierarchical Models for Parametric and
  Semi-Parametric Analyses of Semi-Competing Risks Data}.
\newblock R package version 3.4.

\bibitem[\protect\citeauthoryear{Lee, Rondeau, and Haneuse}{Lee
  et~al.}{2017}]{lee2017accelerated}
Lee, K.~H., V.~Rondeau, and S.~Haneuse (2017).
\newblock Accelerated failure time models for semi-competing risks data in the
  presence of complex censoring.
\newblock {\em Biometrics\/}~{\em 73\/}(4), 1401--1412.

\bibitem[\protect\citeauthoryear{Leffondr{\'e}, Touraine, Helmer, and
  Joly}{Leffondr{\'e} et~al.}{2013}]{leffondre2013}
Leffondr{\'e}, K., C.~Touraine, C.~Helmer, and P.~Joly (2013).
\newblock Interval-censored time-to-event and competing risk with death: is the
  illness-death model more accurate than the cox model?
\newblock {\em International journal of epidemiology\/}~{\em 42\/}(4),
  1177--1186.

\bibitem[\protect\citeauthoryear{Lin, Wei, Yang, and Ying}{Lin
  et~al.}{2000}]{lin-etal-jrss-2000}
Lin, D.~Y., L.~J. Wei, I.~Yang, and Z.~Ying (2000).
\newblock {Semiparametric regression for the mean and rate functions of
  recurrent events}.
\newblock {\em J. Roy. Statist. Soc., ser. B\/}~{\em 62}, 711--730.

\bibitem[\protect\citeauthoryear{Liquet, Timsit, and Rondeau}{Liquet
  et~al.}{2012}]{liquet2012investigating}
Liquet, B., J.-F. Timsit, and V.~Rondeau (2012).
\newblock Investigating hospital heterogeneity with a multi-state frailty
  model: application to nosocomial pneumonia disease in intensive care units.
\newblock {\em BMC Medical Research Methodology\/}~{\em 12\/}(1), 1--14.

\bibitem[\protect\citeauthoryear{{Maja Pohar Perme} and Gerster}{{Maja Pohar
  Perme} and Gerster}{2017}]{pseudo}
{Maja Pohar Perme} and M.~Gerster (\, 2017).
\newblock {\em pseudo: Computes Pseudo-Observations for Modeling}.
\newblock R package version 1.4.3.

\bibitem[\protect\citeauthoryear{Marshall, Garg, Jackson, Holmes, and
  Chase}{Marshall et~al.}{1993}]{marshall1993}
Marshall, G., S.~K. Garg, W.~E. Jackson, D.~L. Holmes, and H.~P. Chase (1993).
\newblock Factors influencing the onset and progression of diabetic retinopathy
  in subjects with insulin-dependent diabetes mellitus.
\newblock {\em Ophthalmology\/}~{\em 100\/}(8), 1133--1139.

\bibitem[\protect\citeauthoryear{Moertel, Fleming, Macdonald, Haller, Laurie,
  Goodman, Ungerleider, Emerson, Tormey, Glick, et~al.}{Moertel
  et~al.}{1990}]{moertel1990levamisole}
Moertel, C.~G., T.~R. Fleming, J.~S. Macdonald, D.~G. Haller, J.~A. Laurie,
  P.~J. Goodman, J.~S. Ungerleider, W.~A. Emerson, D.~C. Tormey, J.~H. Glick,
  et~al. (1990).
\newblock Levamisole and fluorouracil for adjuvant therapy of resected colon
  carcinoma.
\newblock {\em New England Journal of Medicine\/}~{\em 322\/}(6), 352--358.

\bibitem[\protect\citeauthoryear{Munholland and Kalbfleisch}{Munholland and
  Kalbfleisch}{1991}]{munholland1991}
Munholland, P.~L. and J.~D. Kalbfleisch (1991).
\newblock A semi-markov model for insect life history data.
\newblock {\em Biometrics\/}~{\em 47\/}(3), 1117--1126.

\bibitem[\protect\citeauthoryear{O'Keeffe, Tom, and Farewell}{O'Keeffe
  et~al.}{2013}]{okeeffe2013}
O'Keeffe, A.~G., B.~D. Tom, and V.~T. Farewell (2013).
\newblock Mixture distributions in multi-state modelling: some considerations
  in a study of psoriatic arthritis.
\newblock {\em Statistics in Medicine\/}~{\em 32\/}(4), 600--619.

\bibitem[\protect\citeauthoryear{Overgaard}{Overgaard}{2019}]{overgaard-19}
Overgaard, M. (2019).
\newblock {State occupation probabilities in non-Markov models}.
\newblock {\em Math. Meth. Statist.\/}~{\em 28}, 279--290.

\bibitem[\protect\citeauthoryear{Overgaard, Andersen, and Parner}{Overgaard
  et~al.}{2023}]{OAP-stata}
Overgaard, M., P.~K. Andersen, and E.~T. Parner (2023).
\newblock {Pseudo-observations in a multi-state setting}.
\newblock {\em The Stata Journal\/}~{\em 23}, 491--517.

\bibitem[\protect\citeauthoryear{Overgaard, Parner, and Pedersen}{Overgaard
  et~al.}{2017}]{OPP-pseudo}
Overgaard, M., E.~T. Parner, and J.~Pedersen (2017).
\newblock Asymptotic theory of generalized estimating equations based on
  jack-knife pseudo-observations.
\newblock {\em Ann. Statist.\/}~{\em 45}, 1988--2015.

\bibitem[\protect\citeauthoryear{Overgaard, Parner, and Pedersen}{Overgaard
  et~al.}{2019}]{OPP-covdepcens}
Overgaard, M., E.~T. Parner, and J.~Pedersen (2019).
\newblock Pseudo-observations under covariate-dependent censoring.
\newblock {\em J. Statist. Plan. and Inf.\/}~{\em 202}, 112--122.

\bibitem[\protect\citeauthoryear{Parner, Andersen, and Overgaard}{Parner
  et~al.}{2023}]{PAP-IJpseudo}
Parner, E.~T., P.~K. Andersen, and M.~Overgaard (2023).
\newblock {Regression models for censored time-to-event data using
  infinitesimal jack-knife pseudo-observations, with applications to
  left-truncation}.
\newblock {\em Lifetime Data Analysis\/}~{\em 29}, 654–671.

\bibitem[\protect\citeauthoryear{Poynard, Bedossa, Opolon, and {for the
  OBSVIRC, METAVIR, CLINIVIR and DOSVIRC groups}}{Poynard
  et~al.}{1997}]{poynard1997}
Poynard, T., P.~Bedossa, P.~Opolon, and {for the OBSVIRC, METAVIR, CLINIVIR and
  DOSVIRC groups} (1997).
\newblock {Natural history of liver fibrosis progression in patients with
  chronic hepatitis C}.
\newblock {\em Lancet\/}~{\em 349\/}(9055), 825--832.

\bibitem[\protect\citeauthoryear{Putter and Spitoni}{Putter and
  Spitoni}{2018}]{putter2018non}
Putter, H. and C.~Spitoni (2018).
\newblock Non-parametric estimation of transition probabilities in non-markov
  multi-state models: the landmark aalen--johansen estimator.
\newblock {\em Statistical Methods in Medical Research\/}~{\em 27\/}(7),
  2081--2092.

\bibitem[\protect\citeauthoryear{Putter and van Houwelingen}{Putter and van
  Houwelingen}{2015}]{putter2015frailties}
Putter, H. and H.~C. van Houwelingen (2015).
\newblock Frailties in multi-state models: Are they identifiable? do we need
  them?
\newblock {\em Statistical Methods in Medical Research\/}~{\em 24\/}(6),
  675--692.

\bibitem[\protect\citeauthoryear{Reulen}{Reulen}{2022}]{simMSM}
Reulen, H. (\, 2022).
\newblock {\em simMSM: Simulation of Event Histories for Multi-State Models}.
\newblock R package version 1.1.42.

\bibitem[\protect\citeauthoryear{Roimi, Gutman, Somer, Ben~Arie, Calman,
  Bar-Lavie, Gelbshtein, Liverant-Taub, Ziv, Eytan, et~al.}{Roimi
  et~al.}{2021}]{roimi2021development}
Roimi, M., R.~Gutman, J.~Somer, A.~Ben~Arie, I.~Calman, Y.~Bar-Lavie,
  U.~Gelbshtein, S.~Liverant-Taub, A.~Ziv, D.~Eytan, et~al. (2021).
\newblock Development and validation of a machine learning model predicting
  illness trajectory and hospital utilization of covid-19 patients: a
  nationwide study.
\newblock {\em Journal of the American Medical Informatics Association\/}~{\em
  28\/}(6), 1188--1196.

\bibitem[\protect\citeauthoryear{Rondeau, Gonzalez, Mazroui, Mauguen, Diakite,
  Laurent, Lopez, Kr\'{o}l, and Sofeu}{Rondeau et~al.}{2019}]{frailtypack}
Rondeau, V., J.~R. Gonzalez, Y.~Mazroui, A.~Mauguen, A.~Diakite, A.~Laurent,
  M.~Lopez, A.~Kr\'{o}l, and C.~L. Sofeu (2019).
\newblock {\em {frailtypack}: General Frailty Models: Shared, Joint and Nested
  Frailty Models with Prediction; Evaluation of Failure-Time Surrogate
  Endpoints}.
\newblock {R} package version 3.0.3.

\bibitem[\protect\citeauthoryear{Rossman, Keshet, and Gorfine}{Rossman
  et~al.}{2022}]{rossman2022pymsm}
Rossman, H., A.~Keshet, and M.~Gorfine (2022).
\newblock Pymsm: Python package for competing risks and multi-state models for
  survival data.
\newblock {\em Journal of Open Source Software\/}~{\em 7\/}(78), 4566.

\bibitem[\protect\citeauthoryear{Rossman, Meir, Somer, Shilo, Gutman, Ben~Arie,
  Segal, Shalit, and Gorfine}{Rossman et~al.}{2021}]{rossman2021hospital}
Rossman, H., T.~Meir, J.~Somer, S.~Shilo, R.~Gutman, A.~Ben~Arie, E.~Segal,
  U.~Shalit, and M.~Gorfine (2021).
\newblock Hospital load and increased covid-19 related mortality in israel.
\newblock {\em Nature Communications\/}~{\em 12\/}(1), 1904.

\bibitem[\protect\citeauthoryear{Saag, Wagman, Geusens, Adachi, Messina, Emkey,
  Chapurlat, Wang, Pannacciulli, and Lems}{Saag et~al.}{2018}]{saag2018}
Saag, K.~G., R.~B. Wagman, P.~Geusens, J.~D. Adachi, O.~D. Messina, R.~Emkey,
  R.~Chapurlat, A.~Wang, N.~Pannacciulli, and W.~F. Lems (2018).
\newblock Denosumab versus risedronate in glucocorticoid-induced osteoporosis:
  a multicentre, randomised, double-blind, active-controlled, double-dummy,
  non-inferiority study.
\newblock {\em The lancet Diabetes \& Endocrinology\/}~{\em 6\/}(6), 445--454.

\bibitem[\protect\citeauthoryear{Satten}{Satten}{1999}]{satten1999}
Satten, G.~A. (1999).
\newblock Estimating the extent of tracking in interval-censored
  chain-of-events data.
\newblock {\em Biometrics\/}~{\em 55\/}(4), 1228--1231.

\bibitem[\protect\citeauthoryear{Sauerbrei, Abrahamowicz, Altman, le~Cessie,
  Carpenter, and initiative}{Sauerbrei et~al.}{2014}]{sauerbrei2014}
Sauerbrei, W., M.~Abrahamowicz, D.~G. Altman, S.~le~Cessie, J.~Carpenter, and
  S.~initiative (2014).
\newblock Strengthening analytical thinking for observational studies: the
  stratos initiative.
\newblock {\em Statistics in Medicine\/}~{\em 33\/}(30), 5413--5432.

\bibitem[\protect\citeauthoryear{Scheike and Zhang}{Scheike and
  Zhang}{2007}]{ts-mjz-sjs07}
Scheike, T.~H. and M.~Zhang (2007).
\newblock Direct modelling of regression effects for transition probabilities
  in multistate models.
\newblock {\em Scand. J. Statist.\/}~{\em 34}, 17--32.

\bibitem[\protect\citeauthoryear{Scheike, Zhang, and Gerds}{Scheike
  et~al.}{2008}]{directbinomial}
Scheike, T.~H., M.~Zhang, and T.~A. Gerds (2008).
\newblock {Predicting cumulative incidence probability by direct binomial
  regression}.
\newblock {\em Biometrika\/}~{\em 95}, 205--220.

\bibitem[\protect\citeauthoryear{Shih and Louis}{Shih and
  Louis}{1995}]{shih1995inferences}
Shih, J.~H. and T.~A. Louis (1995).
\newblock Inferences on the association parameter in copula models for
  bivariate survival data.
\newblock {\em Biometrics\/}, 1384--1399.

\bibitem[\protect\citeauthoryear{Sood, Petersen, Qualls, Meek,
  Vazquez-Guillamet, Celli, and Tesfaigzi}{Sood
  et~al.}{2016}]{sood2016spirometric}
Sood, A., H.~Petersen, C.~Qualls, P.~M. Meek, R.~Vazquez-Guillamet, B.~R.
  Celli, and Y.~Tesfaigzi (2016).
\newblock Spirometric variability in smokers: transitions in copd diagnosis in
  a five-year longitudinal study.
\newblock {\em Respiratory Research\/}~{\em 17}, 1--10.

\bibitem[\protect\citeauthoryear{Spitoni, Verduijn, and Putter}{Spitoni
  et~al.}{2012}]{spitoni2012estimation}
Spitoni, C., M.~Verduijn, and H.~Putter (2012).
\newblock Estimation and asymptotic theory for transition probabilities in
  markov renewal multi-state models.
\newblock {\em The International Journal of Biostatistics\/}~{\em 8\/}(1).

\bibitem[\protect\citeauthoryear{Sudlow, Gallacher, Allen, Beral, Burton,
  Danesh, Downey, Elliott, Green, Landray, et~al.}{Sudlow
  et~al.}{2015}]{sudlow2015uk}
Sudlow, C., J.~Gallacher, N.~Allen, V.~Beral, P.~Burton, J.~Danesh, P.~Downey,
  P.~Elliott, J.~Green, M.~Landray, et~al. (2015).
\newblock Uk biobank: an open access resource for identifying the causes of a
  wide range of complex diseases of middle and old age.
\newblock {\em PLoS Medicine\/}~{\em 12\/}(3), e1001779.

\bibitem[\protect\citeauthoryear{Sutradhar and Cook}{Sutradhar and
  Cook}{2008}]{sutradhar2008}
Sutradhar, R. and R.~J. Cook (2008).
\newblock Analysis of interval-censored data from clustered multistate
  processes: application to joint damage in psoriatic arthritis.
\newblock {\em Journal of the Royal Statistical Society Series C: Applied
  Statistics\/}~{\em 57\/}(5), 553--566.

\bibitem[\protect\citeauthoryear{Sweeting, Farewell, and De~Angelis}{Sweeting
  et~al.}{2010}]{sweeting2010}
Sweeting, M., V.~Farewell, and D.~De~Angelis (2010).
\newblock Multi-state markov models for disease progression in the presence of
  informative examination times: An application to hepatitis c.
\newblock {\em Statistics in Medicine\/}~{\em 29\/}(11), 1161--1174.

\bibitem[\protect\citeauthoryear{Therneau}{Therneau}{2023}]{therneau2020package}
Therneau, T. (2023).
\newblock A package for survival analysis in r. r package version 3.5-5. 2023.
\newblock {\em URL https://CRAN. R-project. org/package= survival\/}.

\bibitem[\protect\citeauthoryear{Therneau, Crowson, and Atkinson}{Therneau
  et~al.}{2024}]{therneau2024multi}
Therneau, T., C.~Crowson, and E.~Atkinson (2024).
\newblock Multi-state models and competing risks.
\newblock {\em CRAN-R (https://cran. r-project.
  org/web/packages/survival/vignettes/compete. pdf)\/}.

\bibitem[\protect\citeauthoryear{Therneau and Grambsch}{Therneau and
  Grambsch}{2013}]{therneau2013modeling}
Therneau, T. and P.~Grambsch (2013).
\newblock {\em Modeling Survival Data: Extending the Cox Model}.
\newblock Statistics for Biology and Health. Springer New York.

\bibitem[\protect\citeauthoryear{Therneau}{Therneau}{2021}]{survival-package}
Therneau, T.~M. (2021).
\newblock {\em A Package for Survival Analysis in R}.
\newblock R package version 3.2-13.

\bibitem[\protect\citeauthoryear{Titman}{Titman}{2015}]{titman2015}
Titman, A.~C. (2015).
\newblock Transition probability estimates for non-markov multi-state models.
\newblock {\em Biometrics\/}~{\em 71\/}(4), 1034--1041.

\bibitem[\protect\citeauthoryear{Titman and Putter}{Titman and
  Putter}{2020}]{titman2020general}
Titman, A.~C. and H.~Putter (2020, 09).
\newblock General tests of the markov property in multi-state models.
\newblock {\em Biostatistics\/}~{\em 23\/}(2), 380--396.

\bibitem[\protect\citeauthoryear{Titman and Sharples}{Titman and
  Sharples}{2010}]{titman2010}
Titman, A.~C. and L.~D. Sharples (2010).
\newblock Semi-markov models with phase-type sojourn distributions.
\newblock {\em Biometrics\/}~{\em 66\/}(3), 742--752.

\bibitem[\protect\citeauthoryear{Touraine, Gerds, and Joly}{Touraine
  et~al.}{2017}]{touraine2017smoothhazard}
Touraine, C., T.~A. Gerds, and P.~Joly (2017).
\newblock Smoothhazard: an r package for fitting regression models to
  interval-censored observations of illness-death models.
\newblock {\em Journal of Statistical Software\/}~{\em 79}, 1--22.

\bibitem[\protect\citeauthoryear{{van Houwelingen} and Putter}{{van
  Houwelingen} and Putter}{2012}]{HansvHbook}
{van Houwelingen}, H.~C. and H.~Putter (2012).
\newblock {\em Dynamic Prediction in Clinical Survival Analysis}.
\newblock Boca Raton: Chapman and Hall/CRC.

\bibitem[\protect\citeauthoryear{Vaupel, Manton, and Stallard}{Vaupel
  et~al.}{1979}]{vaupel1979impact}
Vaupel, J.~W., K.~G. Manton, and E.~Stallard (1979).
\newblock The impact of heterogeneity in individual frailty on the dynamics of
  mortality.
\newblock {\em Demography\/}~{\em 16\/}(3), 439--454.

\bibitem[\protect\citeauthoryear{Xu, Kalbfleisch, and Tai}{Xu
  et~al.}{2010}]{xu2010statistical}
Xu, J., J.~D. Kalbfleisch, and B.~Tai (2010).
\newblock Statistical analysis of illness--death processes and semicompeting
  risks data.
\newblock {\em Biometrics\/}~{\em 66\/}(3), 716--725.

\bibitem[\protect\citeauthoryear{Yang and Nair}{Yang and Nair}{2011}]{yang2011}
Yang, Y. and V.~N. Nair (2011).
\newblock Parametric inference for time-to-failure in multi-state semi-markov
  models: A comparison of marginal and process approaches.
\newblock {\em Canadian Journal of Statistics\/}~{\em 39\/}(3), 537--555.

\bibitem[\protect\citeauthoryear{Zarghami, Fuh-Ngwa, Claflin, van~der Mei,
  Ponsonby, Broadley, Simpson-Yap, Group, Lucas, Dear, et~al.}{Zarghami
  et~al.}{2024}]{zarghami2024}
Zarghami, A., V.~Fuh-Ngwa, S.~B. Claflin, I.~van~der Mei, A.-L. Ponsonby,
  S.~Broadley, S.~Simpson-Yap, A.~I. Group, R.~Lucas, K.~Dear, et~al. (2024).
\newblock Changes in employment status over time in multiple sclerosis
  following a first episode of central nervous system demyelination, a markov
  multistate model study.
\newblock {\em European Journal of Neurology\/}~{\em 31\/}(1), e16016.

\end{thebibliography}


\newpage

\setcounter{section}{0}
\renewcommand\thesection{S\arabic{section}}

\begin{center}
{\huge  Supplementary Material}    
\end{center}






\section{Code of Section 2.2.1 - The Colon Cancer Study Revisited, I}

\begin{verbatim}
    

palette("Okabe-Ito")

c1 <- subset(colon, etype== 1) # recurrence
c2 <- subset(colon, etype== 2) # death
# There are 5 subjects with recurrence coded on the same day as death.
# Make them recurrent one day earlier than their death (avoid 0 length intervals)
tied.id <- (c1$time == c2$time & c1$status==1 & c2$status==1)
c1$time[tied.id] <- c1$time[tied.id] -1

# Create the counting process dataset
# The variable time1 indicates the start of each interval time2 marks its end. 
# The variable state denotes the state entered at the end of the corresponding interval.
cdata <- tmerge(subset(c2, ,-c(study, time, status, etype)),
                c2, id=id, death= event(time, status),
                options=list(tstart="time1", tstop="time2"))
cdata <- tmerge(cdata, c1, id= id, recur= event(time, status),
                td.recur = tdc(time))

> head(cdata[,c(1:4,13:16,18:21)],5)
  id      rx sex age    time1    time2 death recur                 state age10 male trt
1  1 Lev+5FU   1  43 0.000000 2.650240     0     1                 recur   4.3    1   1
2  1 Lev+5FU   1  43 2.650240 4.164271     1     0 death post-recurrence   4.3    1   1
3  2 Lev+5FU   1  63 0.000000 8.451745     0     0                censor   6.3    1   1
4  3     Obs   0  71 0.000000 1.483915     0     1                 recur   7.1    0   0
5  3     Obs   0  71 1.483915 2.636550     1     0 death post-recurrence   7.1    0   0

# Make the state variable with recurrence and death pre and post recurrence
cdata$state <- with(cdata, factor(recur+ 2*death + 3*td.recur, c(0,1,2,5,3),
                                  c("censor", "recur", "death pre-recurrence", 
                                    "death post-recurrence","censor")))

> table(cdata$state)

               censor                 recur  death pre-recurrence death post-recurrence 
                  475                   468                    38                   414 

cdata$trt <- 1*(cdata$rx == "Lev+5FU")
# a 0/1 treatment variable, 1=Lev+5FU, 0= Observation or Lev
# Output in years gives nicer plot axes
cdata$time1 <- cdata$time1/365.25
cdata$time2 <- cdata$time2/365.25
# Check for consistency
check <- survcheck(Surv(time1, time2, state) ~ 1, cdata, id=id)
check


#Nelson-Aalen (NA) estimator of cumulative transition intensities
na0 <- survfit(Surv(time1, time2, state) ~ 1, cdata, id=id, stype=1, ctype=1) # overall
na1 <- survfit(Surv(time1, time2, state) ~ trt, cdata, id=id, stype=1, ctype=1) # by arm   

# Extract cumulative hazard data from na1 with summary
na1_summary <- summary(na1, cumhaz = TRUE)

na1_df <- data.frame(
  time = na1_summary$time,
  strata = factor(na1_summary$strata, labels = c("Control", "5FU+Lev")),
  cumhaz = na1_summary$cumhaz[,1],       # Only the first transition
  std_err = na1_summary$std.chaz[,1]      # Only the standard error for the first transition
)

# Add confidence intervals for the first transition
na1_df <- na1_df %>%
  mutate(
    lower = pmax(0, cumhaz - 1.96 * std_err),  # 95% confidence interval lower bound
    upper = cumhaz + 1.96 * std_err            # 95% confidence interval upper bound
  )

# Plot2(a)
ggplot(na1_df, aes(x = time, y = cumhaz, color = strata, fill = strata)) +
  geom_line(size = 1) + # Plot the cumulative hazard curve
  geom_ribbon(aes(ymin = lower, ymax = upper), alpha = 0.1) +  # Add shaded area
  labs(x = "Years Since Enrollment", y = expression('NA Estimate of '*Lambda[0][1](t))) +
  scale_color_manual(values = c("5FU+Lev" = "red", "Control" = "blue")) +  
  scale_fill_manual(values = c("5FU+Lev" = "red", "Control" = "blue")) +   
  theme_minimal() +
  theme(
    axis.title.x = element_text(size = 16),  # Change size for x-axis title
    axis.title.y = element_text(size = 16),
    legend.title = element_blank(),      # Remove legend title for simplicity
    legend.position = c(0.1, 0.9),        # Position the legend in the upper left corner
    panel.border = element_rect(color = "black", fill = NA, size = 0.4),
    panel.grid = element_blank()
      ) +
  guides(color = guide_legend(override.aes = list(size = 2)))  # Adjust legend line width

#plot2(b)
na1_df <- data.frame(
  time = na1_summary$time,
  strata = factor(na1_summary$strata, labels = c("Control", "5FU+Lev")),
  cumhaz_1_2 = na1_summary$cumhaz[,2],
  std_err_1_2 = na1_summary$std.chaz[,2],
  cumhaz_2_3 = na1_summary$cumhaz[,3],
  std_err_2_3 = na1_summary$std.chaz[,3]
)

na1_long <- na1_df %>%
  pivot_longer(
    cols = starts_with("cumhaz"),
    names_to = "transition",
    values_to = "cumhaz"
  ) %>%
  mutate(
    std_err = case_when(
      transition == "cumhaz_1_2" ~ std_err_1_2,
      transition == "cumhaz_2_3" ~ std_err_2_3
    ),
    lower =  pmax(0, cumhaz - 1.96 * std_err),  # Calculate lower CI
    upper = cumhaz + 1.96 * std_err    # Calculate upper CI
  )

ggplot(na1_long, aes(x = time, y = cumhaz, color = strata, fill = strata)) +
  geom_line(size = 1, aes(linetype = transition, group = interaction(strata, transition))) +  
  geom_ribbon(aes(ymin = lower, ymax = upper, 
    group = interaction(strata, transition)), alpha = 0.1) +  
  labs(x = "Years Since Enrollment", y = expression('NA Estimate of '*Lambda[j][2](t))) +
  
  # Set custom colors and fills for strata
  scale_color_manual(
    values = c("5FU+Lev" = "red", "Control" = "blue"),
    labels = c("5FU+Lev" = "5FU+Lev", "Control" = "Control")  
  ) +
  scale_fill_manual(values = c("5FU+Lev" = "red", "Control" = "blue")) +
  
  # Define custom linetype labels using TeX for mathematical notation
  scale_linetype_manual(
    values = c("solid", "dashed"),
    labels = c(TeX("$\\Lambda_{\\0\\2}$"), TeX("$\\Lambda_{12\\'}$"))  
  ) +
  # Adjust theme and legend
  theme_minimal() +
  theme(
    axis.title.x = element_text(size = 16),  # Change size for x-axis title
    axis.title.y = element_text(size = 16),
    legend.title = element_blank(),           # Remove legend title
    legend.position = c(0.1, 0.9),            # Position legend in the upper left corner
    legend.text = element_text(size = 14),
    panel.border = element_rect(color = "black", fill = NA, size = 0.4),
    panel.grid = element_blank()
      ) +
  # Override aesthetics for clearer legend display
  guides(
    color = guide_legend(override.aes = list(size = 1.2)),
    linetype = guide_legend(override.aes = list(size = 1.2))
  )


# Plots 2(c) and 2(d)
#AJ for CIF
plot(na1$time[1:795],(na1$pstate[,"recur"]+na1$pstate[,"death post-recurrence"])[1:795],
     type = 's',
     lwd=2, lty=1, col=c("blue"),
     xlab="Years since enrollment", ylab=expression('AJ Estimate of '*F[1](t)))
lines(na1$time[796:1193],(na1$pstate[,"recur"]+na1$pstate[,"death post-recurrence"])[796:1193],
     type = 's',
     lwd=2, lty=1, col=c("red"))
legend(-0.3, 0.6, c("5FU+Lev", "Control"),
       lwd=2, lty=1, col=c("red","blue"), bty='n')

plot(na1, noplot=c("(s0)","recur"),
     lwd=2, lty=c(1,1,2,2), col=c("blue","red", "blue","red"),
     xlab="Years since enrollment", 
     ylab=expression('AJ Estimate of '*F[2](t)* ' and ' *F[2*"'"](t)))
legend(-0.2, 0.55, c(expression(F[2](t)*" 5FU+Lev"), 
                     expression(F[2*"'"](t)*' 5FU+Lev'),
                     expression(F[2](t)*' Control'), 
                     expression(F[2*"'"](t)*' Control')),
       lwd=2, lty=1:2, col=c("red", "red", "blue", "blue"), bty='n')

%\end{minted}    
%\end{spacing}
\end{verbatim}

\section{Code of Section 2.4.1 - The Colon Cancer Study Revisited, II}

\begin{verbatim}

> table(cdata$extent)
   1    2    3    4 
  26  140 1157   72 
  
> table(cdata$extent,cdata$state)   
    censor recur death pre-recurrence death post-recurrence
  1     17     5                    1                     3
  2     70    34                    5                    31
  3    374   400                   28                   355
  4     14    29                    4                    25

  > cfit2New <- coxph(Surv(time1, time2, state) ~ trt + extent01 + node4
+                   ,data= cdata, id=id, robust = FALSE)
> cfit2New
Call:
coxph(formula = Surv(time1, time2, state) ~ trt + extent01 + 
    node4, data = cdata, robust = FALSE, id = id)

          
1:2            coef exp(coef) se(coef)      z        p
  trt      -0.50584   0.60300  0.10628 -4.759 1.94e-06
  extent01  0.64909   1.91380  0.16803  3.863 0.000112
  node4     0.84501   2.32799  0.09594  8.807  < 2e-16

          
1:3          coef exp(coef) se(coef)     z     p
  trt      0.0346    1.0352   0.3331 0.104 0.917
  extent01 0.1084    1.1145   0.4488 0.242 0.809
  node4    0.4864    1.6265   0.3733 1.303 0.193

          
2:4          coef exp(coef) se(coef)     z        p
  trt      0.2346    1.2645   0.1126 2.083 0.037231
  extent01 0.3040    1.3552   0.1796 1.692 0.090642
  node4    0.3792    1.4610   0.1031 3.678 0.000235

 States:  1= (s0), 2= recur, 3= death pre-recurrence, 4= death post-recurrence 

Likelihood ratio test=143.7  on 9 df, p=< 2.2e-16
n= 1395, unique id= 929, number of events= 920 
> # 95% confidence intervals of HR
> temp <- summary(cfit2New)$conf.int[ , c(1,3,4)]
> temp
             exp(coef) lower .95 upper .95
trt_1:2      0.6029969 0.4896068 0.7426475
extent01_1:2 1.9137966 1.3767796 2.6602787
node4_1:2    2.3279944 1.9289178 2.8096367
trt_1:3      1.0352090 0.5388494 1.9887888
extent01_1:3 1.1144967 0.4624284 2.6860437
node4_1:3    1.6264570 0.7824650 3.3808057
trt_2:4      1.2644642 1.0139800 1.5768257
extent01_2:4 1.3552118 0.9530066 1.9271629
node4_2:4    1.4610437 1.1937605 1.7881717

# generate the forest plot - Figure 3
opar <- par(mar=c(5,7, 1, 1))  # more room for y axis labels
yy <- c(outer(3:1, c(8, 0, 4), "+"))
plot(c(.5,4), range(yy), log='x',  type="n", 
     xlab="Hazard Ratio", ylab="", yaxt= 'n')
axis(2, at=yy, label=rep(c("Lev+5FU", "Extent 3 or 4", "Nodes > 4") ,3),
     las=2)

points(temp[,1], yy, pch=19)
segments(temp[,2], yy, temp[,3], yy, lwd=1.5)

abline(h=c(4,8), lwd=2, lty=2, col="red")
abline(v=1, lty=3)
text(c(2.5, 2.5, 2.5), c(11, 6.5, 3.5), c("Entry to Recurrence", "Recurrence to Death", 
                                     "Entry to Death"))
par(opar)


\end{minted}    

\section{Code of Section 4.1 - The Psoriatic Arthritis Data Revisited}

\begin{minted}{R}
baseline_df <- psor %>%
  group_by(ptnum) %>%
  filter(months == min(months)) %>%
  select(ptnum, hieffusn_baseline = hieffusn, ollwsdrt_baseline = ollwsdrt)

baseline_df$ollwsdrt_baseline <- ifelse(baseline_df$ollwsdrt_baseline==0,1,0)

merged_df <- psor %>%
  left_join(baseline_df, by = "ptnum")

cutpoints <- c(-Inf, 5, 10, 20, Inf)  

# Create a new categorical variable for the intervals
merged_df <- merged_df %>%
  mutate(
    time_interval = cut(months, breaks = cutpoints, 
                        labels = c("1", "2", "3", "4"),
                        include.lowest = TRUE)
  )

> head(merged_df[,c(1:3,6:8)],9)
  ptnum  months state hieffusn_baseline ollwsdrt_baseline time_interval
1     1  6.4606     1                 0                 0             2
2     1 17.0780     1                 0                 0             3
3     2 26.3217     1                 0                 1             4
4     2 29.4839     3                 0                 1             4
5     2 30.5763     4                 0                 1             4
6     5  0.0753     1                 0                 1             1
7     5 12.8939     1                 0                 1             3
8     5 13.3730     2                 0                 1             3
9     5 14.3504     2                 0                 1             3

psor.q <- rbind(c(0,0.1,0,0),
                c(0,0,0.1,0),
                c(0,0,0,0.1),
                c(0,0,0,0))
psorpci.msm <- msm(state ~ months, subject = ptnum,
                   data = merged_df, 
                   qmatrix = psor.q, 
                   covariates = ~ hieffusn_baseline + ollwsdrt_baseline,
                   gen.inits=TRUE,
                   pci = c(5,10, 20),
                   control = list(REPORT=1,trace=2))
HR = hazard.msm(psorpci.msm)
> HR
$hieffusn_baseline
                        HR         L        U
State 1 - State 2 2.100252 0.9595722 4.596899
State 2 - State 3 1.709818 0.9548091 3.061846
State 3 - State 4 1.358133 0.7385758 2.497407

$ollwsdrt_baseline
                         HR         L        U
State 1 - State 2 1.2704387 0.7372401 2.189266
State 2 - State 3 2.1677660 1.2500937 3.759086
State 3 - State 4 0.6981905 0.3422453 1.424329

$`timeperiod[5,10)`
                         HR         L        U
State 1 - State 2 0.7181648 0.3868209 1.333332
State 2 - State 3 0.8283696 0.3936377 1.743218
State 3 - State 4 1.3251053 0.4340788 4.045128

$`timeperiod[10,20)`
                         HR         L         U
State 1 - State 2 0.4195343 0.1874067 0.9391822
State 2 - State 3 0.7961147 0.4064397 1.5593917
State 3 - State 4 1.1737629 0.3810700 3.6153969

$`timeperiod[20,Inf)`
                        HR         L        U
State 1 - State 2 1.565952 0.7626200 3.215501
State 2 - State 3 1.126529 0.5017957 2.529052
State 3 - State 4 1.372552 0.4309212 4.371793


t = seq(0, 60, by = 0.1)
p02 = c(); p03 = c(); p04 = c()
for(i in t){
  p = pmatrix.msm(psorpci.msm, t=i)
  p02 = append(p02, p[1,2])
  p03 = append(p03, p[1,3])
  p04 = append(p04, p[1,4])
}

plot(t, p04, type="l", lwd = 2, col = "blue", xlab = "Time from Diagnosis", 
     ylab = "Transition Probability")
lines(t, p03, lwd = 2, col = "green")
lines(t, p02, lwd = 2, col = "red")
legend(-0.5, 1, c(expression(P[0][1](t)), 
                  expression(P[0][2](t)), 
                  expression(P[0][3](t))),
       lwd=2, lty=1, col=c("red", "green", "blue"), bty='n')


# Figure 4(a)
first5 = psor[psor$ptnum %in% unique(psor$ptnum)[1:5], 1:3]

plot(0, ylim = c(0.7, 5.3), xlim = c(0, 32), xaxt = "n", yaxt = "n", ylab = "Patient ID", 
     xlab = "Years Since Disease Onset", tck = 0, las = 1)
axis(1, at = seq(0, 32, 5))

patient_id = unique(first5$ptnum)

# Create a mapping from patient index to letters
patient_labels = c("A", "B",  "C", "D", "E")

j = 0
for (i in patient_id) {
  id_data = first5[first5$ptnum == i,]
  j = j + 1
  lines(c(id_data$months[1], tail(id_data$months, n=1)), 
        c(j, j), lty = 1)
  points(id_data$months, rep(j, length(id_data$months)), 
         pch = rep("|", length(id_data$months)))
}

# Replace y-axis numbers with letters
axis(2, at = 1:5, labels = patient_labels, las = 2)


# Figure 4(b)
# TransProb
t = seq(0, 30, by = 0.1)
p02 = c()
p03 = c()
p04 = c()
for(i in t){
  p = pmatrix.msm(psorpci.msm, t=i)
  p02 = append(p02, p[1,2])
  p03 = append(p03, p[1,3])
  p04 = append(p04, p[1,4])
}

plot(t, p04, type="l", lwd = 2, col = "blue", xlab = "Years Since Disease Onset", 
     ylab = "Transition Probability")
lines(t, p03, lwd = 2, col = "green")
lines(t, p02, lwd = 2, col = "red")
legend(-0.5, 0.75, c(expression(P[0][1](t)), 
                  expression(P[0][2](t)), 
                  expression(P[0][3](t))),
       lwd=2, lty=1, col=c("red", "green", "blue"), bty='n')
\end{verbatim}    

\section{Codes of Section 5.1.3 - The Rotterdam Tumor Bank Data Revisited.}

All the codes used for Table 3 are available at \url{https://github.com/LeaKats/semicompAFT}. Here we provide also the two codes based on the package \texttt{SemiCompRisks}. The codes of the marginalized Cox and multiplicative AFT approach are too long to be included here and can be found in the above GitHub address.

\subsection{Conditional frailty with Cox-type model}

\begin{verbatim}
#Load packages
suppressPackageStartupMessages(library(SemiCompRisks))
suppressPackageStartupMessages(library(survival))
suppressPackageStartupMessages(library(readxl))

###### Load data ######
data <- as.data.frame(read_excel("breast.xlsx"))
data<-data[data$nodes>0,]

data$size20_50<-ifelse(data$size=="20-50",1,0)
data$size50<-ifelse(data$size==">50",1,0)
data$grade3<-ifelse(data$grade==3,1,0)
data$log_pgr<-log(1+data$pgr)
data$log_er<-log(1+data$er)
data$age_10<-data$age/10
data$log_nodes<-log(data$nodes)

n<-dim(data)[1]

data$rtime_years<-data$rtime/365.25
data$rtime_10<-data$rtime/10
data$rtime_years_10<-data$rtime/365.25/10
data$dtime_10<-data$dtime/10
data$dtime_years<-data$dtime/365.25
#dealing with cases when recur time=death time (added 0.5 day):
data$dtime_years[(data$recur==1)&(data$rtime==data$dtime)]<-
  data$dtime_years[(data$recur==1)&(data$rtime==data$dtime)]+0.5/365.25
data$age_at_relapse_10<-data$age_10+data$rtime_years_10

set.seed(1)

form <- Formula(rtime + recur | dtime + death ~ 
            age_10+log_nodes+log_er+log_pgr+meno+size20_50+size50+hormon+chemo+grade3 |  
            age_10+log_nodes+log_er+log_pgr+meno+size20_50+size50+hormon+chemo+grade3 | 
            age_at_relapse_10+log_nodes+log_er+log_pgr+meno+size20_50+size50+hormon+chemo+grade3)

#####################
## Hyperparameters ##
#####################

## Subject-specific frailty variance component
##  - prior parameters for 1/theta
##
## Subject-specific frailty variance component
##  - prior parameters for 1/theta
##
theta.ab <- c(0.7, 0.7)

## Weibull baseline hazard function: alphas, kappas
##
WB.ab1 <- c(0.5, 0.01) # prior parameters for alpha1
WB.ab2 <- c(0.5, 0.01) # prior parameters for alpha2
WB.ab3 <- c(0.5, 0.01) # prior parameters for alpha3
##
WB.cd1 <- c(0.5, 0.05) # prior parameters for kappa1
WB.cd2 <- c(0.5, 0.05) # prior parameters for kappa2
WB.cd3 <- c(0.5, 0.05) # prior parameters for kappa3

## PEM baseline hazard function
##
PEM.ab1 <- c(0.7, 0.7) # prior parameters for 1/sigma_1^2
PEM.ab2 <- c(0.7, 0.7) # prior parameters for 1/sigma_2^2
PEM.ab3 <- c(0.7, 0.7) # prior parameters for 1/sigma_3^2
##
PEM.alpha1 <- 10 # prior parameters for K1
PEM.alpha2 <- 10 # prior parameters for K2
PEM.alpha3 <- 10 # prior parameters for K3

## MVN cluster-specific random effects
##
Psi_v <- diag(1, 3)
rho_v <- 100

## DPM cluster-specific random effects
##
Psi0  <- diag(1, 3)
rho0  <- 10
aTau  <- 1.5
bTau  <- 0.0125

##
hyperParams <- list(theta=theta.ab,
                    WB=list(WB.ab1=WB.ab1, WB.ab2=WB.ab2, WB.ab3=WB.ab3,
                            WB.cd1=WB.cd1, WB.cd2=WB.cd2, WB.cd3=WB.cd3),
                    PEM=list(PEM.ab1=PEM.ab1, PEM.ab2=PEM.ab2, PEM.ab3=PEM.ab3,
                             PEM.alpha1=PEM.alpha1, PEM.alpha2=PEM.alpha2, PEM.alpha3=PEM.alpha3),
                    MVN=list(Psi_v=Psi_v, rho_v=rho_v),
                    DPM=list(Psi0=Psi0, rho0=rho0, aTau=aTau, bTau=bTau))

###################
## MCMC SETTINGS ##
###################

## Setting for the overall run
##
numReps    <- 20000000
thin       <- 10000
burninPerc <- 0.5

## Settings for storage
##
nGam_save <- 0
storeV    <- rep(TRUE, 3)

## Tuning parameters for specific updates
##
##  - those common to all models
mhProp_theta_var  <- 0.05
mhProp_Vg_var     <- c(0.05, 0.05, 0.05)
##
## - those specific to the Weibull specification of the baseline hazard functions
mhProp_alphag_var <- c(0.01, 0.01, 0.01)
##
## - those specific to the PEM specification of the baseline hazard functions
Cg        <- c(0.2, 0.2, 0.2)
delPertg  <- c(0.5, 0.5, 0.5)
rj.scheme <- 1
Kg_max    <- c(50, 50, 50)
sg_max    <- c(max(data$rtime[data$recur == 1]),
               max(data$dtime[data$recur == 0 & data$death == 1]),
               max(data$dtime[data$recur == 1 & data$death == 1]))

time_lambda1 <- seq(1, sg_max[1], 1)
time_lambda2 <- seq(1, sg_max[2], 1)
time_lambda3 <- seq(1, sg_max[3], 1)               

##
mcmc.WB  <- list(run=list(numReps=numReps, thin=thin, burninPerc=burninPerc),
                 storage=list(nGam_save=nGam_save, storeV=storeV),
                 tuning=list(mhProp_theta_var=mhProp_theta_var,
                             mhProp_Vg_var=mhProp_Vg_var, mhProp_alphag_var=mhProp_alphag_var))

##
mcmc.PEM <- list(run=list(numReps=numReps, thin=thin, burninPerc=burninPerc),
                 storage=list(nGam_save=nGam_save, storeV=storeV),
                 tuning=list(mhProp_theta_var=mhProp_theta_var,
                             mhProp_Vg_var=mhProp_Vg_var, Cg=Cg, delPertg=delPertg,
                             rj.scheme=rj.scheme, Kg_max=Kg_max,
                             time_lambda1=time_lambda1, time_lambda2=time_lambda2,
                             time_lambda3=time_lambda3))

#####################
## Starting Values ##
#####################

##
Sigma_V <- diag(0.1, 3)
Sigma_V[1,2] <- Sigma_V[2,1] <- -0.05
Sigma_V[1,3] <- Sigma_V[3,1] <- -0.06
Sigma_V[2,3] <- Sigma_V[3,2] <- 0.07


#############
## Data analysis: PEM-DPM ##
#PEM: non-parametric mixture of piecewise exponential models 
#DPM: non-parametric Dirichlet process mixture of multivariate normals 
#############

##
myModel <- c("Markov", "PEM")
myPath  <- "Output/02-Results-PEM/"

startValues      <- initiate.startValues_HReg(form, data, model=myModel, nChain=2)

##
fit_PEM <- BayesID_HReg(form, data, id=NULL, model=myModel,
                        hyperParams, startValues, mcmc.PEM, path=myPath)

fit_PEM
summ.fit_PEM <- summary(fit_PEM); names(summ.fit_PEM)
summ.fit_PEM

\end{minted}

\subsection{The AFT additive-frailty model}

\begin{minted}{R}
library(SemiCompRisks)
library(Formula)
library(readxl)

#read data
data <- as.data.frame(read_excel("breast.xlsx"))
data<-data[data$nodes>0,]



#define variables
data$age_10<-data$age/10
data$size20_50<-ifelse(data$size=="20-50",1,0)
data$size50<-ifelse(data$size==">50",1,0)
data$log_nodes<-log(data$nodes)
data$log_pgr<-log(1+data$pgr)
data$log_er<-log(1+data$er)
data$grade3<-ifelse(data$grade==3,1,0)
data$age_since_relapse_10<-data$age_10+data$rtime/10

#define the times used in the format of package SemiCompRisks (in years)
data$y1L<-data$y1U<-data$rtime/365.25
data$y1U[which(data$recur == 0)] <- Inf
data$y2L<-data$y2U<-data$dtime/365.25
data$y2U[which(data$death == 0)] <- Inf
data$LT <- rep(0, dim(data)[1])


#define the formula that includes all transitions
form <- Formula(LT | y1L + y1U | y2L + y2U  ~ 
    age_10+meno+size20_50+size50+log_nodes+log_er+log_pgr+hormon+chemo+grade3 | 
    age_10+meno+size20_50+size50+log_nodes+log_er+log_pgr+hormon+chemo+grade3 | 
    age_since_relapse_10+meno+size20_50+size50+log_nodes+log_er+log_pgr+hormon+chemo+grade3)

#####################
## Define the Hyperparameters 
#####################

## Subject-specific random effects variance component
##
theta.ab <- c(0.5, 0.05)

## log-Normal model
##
LN.ab1 <- c(0.3, 0.3)
LN.ab2 <- c(0.3, 0.3)
LN.ab3 <- c(0.3, 0.3)

## DPM model
##
DPM.mu1 <- log(12)
DPM.mu2 <- log(12)
DPM.mu3 <- log(12)

DPM.sigSq1 <- 100
DPM.sigSq2 <- 100
DPM.sigSq3 <- 100

DPM.ab1 <-  c(2, 1)
DPM.ab2 <-  c(2, 1)
DPM.ab3 <-  c(2, 1)

Tau.ab1 <- c(1.5, 0.0125)
Tau.ab2 <- c(1.5, 0.0125)
Tau.ab3 <- c(1.5, 0.0125)

##
hyperParams <- list(theta=theta.ab,
                    LN=list(LN.ab1=LN.ab1, LN.ab2=LN.ab2, LN.ab3=LN.ab3),
                    DPM=list(DPM.mu1=DPM.mu1, DPM.mu2=DPM.mu2, 
                    DPM.mu3=DPM.mu3, 
                    DPM.sigSq1=DPM.sigSq1,
                    DPM.sigSq2=DPM.sigSq2,
                    DPM.sigSq3=DPM.sigSq3,
                    DPM.ab1=DPM.ab1, 
                    DPM.ab2=DPM.ab2,
                    DPM.ab3=DPM.ab3, 
                    Tau.ab1=Tau.ab1, 
                    Tau.ab2=Tau.ab2, 
                    Tau.ab3=Tau.ab3))

###################
## Define the MCMC SETTINGS ##
###################

## Setting for the overall run
##
numReps    <- 1000000
thin       <- 1000
burninPerc <- 0.5

## Setting for storage
##
nGam_save <- 10
nY1_save <- 10
nY2_save <- 10
nY1.NA_save <- 10

## Tuning parameters for specific updates
##
##  - those common to all models
betag.prop.var	<- c(0.01,0.01,0.01)
mug.prop.var	<- c(0.1,0.1,0.1)
zetag.prop.var	<- c(0.1,0.1,0.1)
gamma.prop.var	<- 0.01

##
mcmcParams <- list(run=list(numReps=numReps, thin=thin, burninPerc=burninPerc),
    storage=list(nGam_save=nGam_save, 
    nY1_save=nY1_save, nY2_save=nY2_save, 
    nY1.NA_save=nY1.NA_save),
    tuning=list(betag.prop.var=betag.prop.var, 
    mug.prop.var=mug.prop.var,
    zetag.prop.var=zetag.prop.var,
    gamma.prop.var=gamma.prop.var))

############
## Analysis 
############

#########
## DPM (Semiparametric Dirichlet process mixture model)##
#########

##
myModel <- "DPM"
myPath  <- paste0("Output/02-Results-DPM/",numReps,"/")

#initiate start values
startValues      <- initiate.startValues_AFT(form, data, model=myModel, nChain=3)

#fit
fit_DPM <- BayesID_AFT(form, data, model=myModel, hyperParams,
                       startValues, mcmcParams, path=myPath)

#summary
summ.fit_DPM <- summary(fit_DPM); names(summ.fit_DPM)
summ.fit_DPM

write.csv(summ.fit_DPM$theta,file=paste0(myPath,r,"_","theta_DPM.csv"))
write.csv(summ.fit_DPM$coef,file=paste0(myPath,r,"_","coef_DPM.csv"))
write.csv(summ.fit_DPM$psrf,file=paste0(myPath,r,"_","psrf_DPM.csv"))
write.csv(summ.fit_DPM$h0,file=paste0(myPath,r,"_","h0_DPM.csv"))
write.csv(summ.fit_DPM$setup[6:11],file=paste0(myPath,r,"_","setup_DPM.csv"))
\end{minted}


\end{verbatim}

\end{document}